%% file: TCAM_arxiv.tex
\documentclass[preprint,authoryear,12pt]{elsarticle}

\journal{Atmospheric Environment}

\usepackage{natbib}
\usepackage{amsmath}
\usepackage{graphicx}
\usepackage{amsfonts}
\usepackage{amssymb}
\usepackage{lineno}

\include{command} % at home
\begin{document}

\begin{frontmatter}
%--------------------------------------------------------------------------------------------------
\title{Comparing mesoscale chemistry-transport model and remote-sensed Aerosol Optical Depth}

\author[BS]{C. Carnevale} 
\author[BS]{G. Mannarini \corref{cor1} \fnref{fn1}} \ead{G.Mannarini @ gmail.com}
\author[BS]{E. Pisoni}
\author[BS]{M. Volta} 

\cortext[cor1]{Corresponding author. Tel:   +39-0832-298 720; Fax:  +39-0832-298 716}
\address[BS]{Dipartimento di Ingegneria dell'Informazione, Universit\agr degli Studi di Brescia via Branze 38, I-25123 ITALY}
%\address[LE]{CNR-ISAC u.o.s. di Lecce, s.p. Lecce-Monteroni km 1.2, I-73100 Lecce ITALY}
\fntext[fn1]{Present address: CNR-ISAC u.o.s. di Lecce, s.p. Lecce-Monteroni km 1.2, I-73100 Lecce ITALY}

%\cortext[cor1]{G.Mannarini@gmail.com 
% 
%\fntext[fn1]{corresponding author}

\begin{abstract}
A comparison of modeled and observed Aerosol Optical Depth (AOD) is presented. 3D Eulerian multiphase chemistry-transport model TCAM is employed for simulating AOD at mesoscale. MODIS satellite sensor and AERONET photometer AOD are used for comparing spatial patterns and temporal timeseries. 

TCAM simulations for year 2004 over a domain containing Po-Valley and nearly whole Northern Italy are employed.

For the computation of AOD, a configuration of external mixing of the chemical species is considered. Furthermore, a parametrization of the effect of moisture affecting both aerosol size and composition is used.

An analysis of the contributions of the granulometric classes to the extinction coefficient reveals the dominant role of the  inorganic compounds of submicron size. For the analysis of spatial patterns, summer and winter case study are considered. TCAM AOD  reproduces spatial patterns similar to those retrieved from space, but AOD values are generally smaller by an order of magnitude. However, accounting also for moisture,  TCAM AOD significantly increases to values of the same magnitude of the observed ones. The temporal performance of model AOD is tested in correspondence of AERONET site "Venise" and some temporal structures are reproduced. 

The results suggest encouraging perspectives in view of satellite AOD assimilation also at the mesoscale and not only at the global and regional scale as already tested in the literature.
\end{abstract}

\begin{keyword} 
Aerosol Optical Depth \sep Chemistry-Transport Model \sep TCAM \sep MODIS \sep Po-Valley \sep remote-sensing
\end{keyword} 
%--------------------------------------------------------------------------------------------------
\end{frontmatter}

%\linenumbers
\section{Introduction}

Chemistry-Transport Models (CTMs) often are the sole valuable tool for investigating the impact of pollutant emissions and air quality strategies at the mesoscale. This is due to the nonlinear nature of the process of pollutant chemical transformation, diffusion and advection \cite{scha04}. 
CTMs provide both gas and aerosol phase state variables, which either exactly  represent the same chemical quantity measured at the monitoring sites (such as O$_3$ or NO$_x$) or can be aggregated for computing monitored quantities (such as PM10). While CTM performance in reproducing observed gas phase quantities is fair, PM10 simulation skill is generally poor \cite{vaut07}. Among the reasons for CTM weak  performance, it is possible to list:   uncertainties in the emission inventories \cite{henz09}, not detailed representation of aerosol phase chemistry, in particular the formation of secondary aerosol \cite{stai08}, as well as difficulties in comparing  point-like experimental information from monitoring sites to spatially averaged values from models \cite{vaut07}. As long as these issues remain unsolved, CTM aerosol simulation and prediction power relies on data assimilation, which makes use of auxiliary observations (e.g.: air quality stations, remote sensing data) in order to reduce the uncertainties in input data, such as initial conditions or boundary conditions \cite{tomb09}. 

Spatially and temporally homogeneous observations on the mesoscale can be provided by means of optical instruments for remote sensing of aerosols. Aerosol Optical Depth (AOD), which is the column integrated extinction coefficient, represents the optical effect of the total aerosol load in the atmosphere. There are both  ground based stations as well as airborne measurements of AOD. 

An AOD output has been computed from various CTMs, both 
at the global \cite{morc09, bene09}, regional \cite{koel06,adhi07,choi09}, and mesoscale   \cite{hodz04,  deme07,  chap09}.
 This brief survey is deliberately restricted to models run at the mesoscale. 

 CHIMERE model was run over the Paris greater metropolitan area with a horizontal spatial resolution of 6 km and 20 vertical layers up to the pressure level of 500 hPa \cite{hodz04}. This model employes a sectional approach with 6 bins between 10 nm and 40 $\mu$m. The model delivers both a dry and a wet particle radius, i.e. a radius accounting for the hygroscopic growth. The wet radius is obtained from a parametrization which depends on relative humidity (RH) only.  Vertical profiles of the aerosol backscattering coefficient were computed and compared to Lidar measurements. The average complex refractive index and the Single Scattering Albedo were compared to Sun photometer measurements. It was found that the aerosol load is typically underestimated by the model and that a prevalence of submicron sized particles is simulated.

In another mesoscale modeling exercise, AOD was computed by TAPOM model over a  portion of the Po-Valley (Italy) \cite{deme07}. Horizontal spatial resolution down to 5 km  and 21 vertical layers up to 6000 m a.s.l  were employed. The model considers a sectional approach with 4 fixed size bins ranging from 0 to 10 $\mu$m.  The activity coefficients of atmospheric aerosols are assumed to be in equilibrium with the environment, governed by RH. Model AOD was compared to both MODIS and MISR satellite observations as well as to AERONET and CHARM Sun photometers data. The model was found to underestimate observations.

Also the WRF-chem model has been developed for computing aerosol chemistry and aerosol optics \cite{fast05, chap09}. A domain comprising western Pennsylvania was modeled with 2 km resolution. Vertically, the  domain extended 57 nodes in the vertical, from the surface to 100 hPa, with finer resolution near the surface. The model was run with 8 dimensional sections, ranging from 20  nm to 5 $\mu$m radius \cite{chap09}.  Thanks to a specific module for calculating the activity coefficients of various electrolytes in multicomponent aqueous atmospheric aerosols and a solid-liquid equilibrium solver, the model is able to describe aerosol water uptake. However, particle transfer among sections as a consequence of water uptake is forbidden. Model AOD was compared to radiometer observations capturing the diurnal  cycle in sequence of a few days, although predicted AOD on a day after precipitation events  is clearly too high.

In the present paper, an optical module is presented which works on top of the aerosol simulations from  TCAM (Transport and Chemical Aerosol Model). TCAM  is a multiphase three-dimensional Eulerian model. A peculiarity of the model is that aerosol particles are described by 10 granulometric classes dynamically set by the model, allowing a detailed description of the chemical reactions affecting particle size \cite{carn06}. The domain of investigation is Po-Valley, in Italy.  Northern Italy is a densely inhabited and industrialized area, where high anthropogenic emissions and frequent stagnating meteorological conditions regularly cause high PM10 levels. For the computation of AOD, an external mixing approach of the aerosol species has been chosen. Hygroscopicity is accounted for in a parametric way. For the sake of making an homogeneous comparison with satellite AOD products, a cloud mask is applied to the model outputs. At the AERONET site "Venise", both model and satellite AOD are compared to the reference value from the photometers. 

The paper contains a detailed description of the optical module for the computation of AOD from TCAM aerosol fields \Sect{sec:modelAOD} and a case study, including comparisons to observations, which is presented in \Sect{sec:results}. A combined Discussion and Conclusions section is also provided \Sect{sec:conclusion}.

\section{TCAM}\label{sec:modelAOD}

\subsection{Aerosol Module}\label{sec:aer_module}
TCAM is a part of the Gas Aerosol Modeling Evaluation System (GAMES) \cite{volt06} which also includes the meteorological pre-processor PROMETEO and the emission processor POEM-PM \cite{carn06}. 

Within TCAM the chemical and physical dynamics of aerosol particles of different sizes is modeled. The aerosol module considers 10 size bins (usually ranging between 10nm and 30$\mu$m), whose bounds can vary during the simulation, according to the gas-to-particle and growth phenomena, and according to the concentration of each compound. This detailed description of the granulometric classes within TCAM  is an important difference with respect to other CTMs.

The aerosol module implemented in TCAM is coupled with the COCOH97 gas phase chemical mechanism. TCAM describes the most relevant aerosol processes:  the condensation, the evaporation, the nucleation of H$_2$SO$_4$  and the aqueous
oxidation of SO$_2$.

Water is assumed to be always in equilibrium between the gas and
the aerosol phases, i.e., water activity of an aqueous aerosol is
equal to the ambient RH. Thus, water is not a transported quantity: it just mirrors the meteorological input. This assumption is also made by other model
developers \citep{deme07, wexl94}. 

More technical details about the model and its performances may be found in \cite{carn08}.

\subsection{Optical module}\label{sec:opt_module}
The optical module takes as an input the aerosol mass concentrations resulting from TCAM simulations and delivers the associated AOD.
The approach is based on the computation of the  extinction coefficient $\alpha$, from which AOD is obtained after integration over the optical path,
\be\label{AOD_def}
 AOD = \int_{h_1}^{h_{11}}\, \alpha(z) \, dz 
\ee
Aerosol particles are considered as externally mixed. Thereafter, $\alpha$ results from linear superposition of the contributions $\alpha^{(s)}$ of different aerosol species. Furthermore, each of the $n_K$ model size classes  has to be added up for obtaining $\alpha^{(s)}$. This leads to the double summation:
\be\label{alpha_as_a_sum}
 \alpha = \sum_{s=1}^{n_S} \sum_{k=1}^{n_K} \alpha_k^{(s)}
\ee
Individual addends in the extinction coefficient are given by the number concentration times the extinction cross section \cite{BH83},
\be\label{extcoeff_def}
\alpha_k^{(s)} = N_k^{(s)}  \, \sigma^{(s)}_\mathrm{ext}(r_k )
\ee
The extinction cross section $\sigma^{(s)}_\mathrm{ext}$ is the geometrical cross section of the particles (assumed to be spherical) multiplied by the extinction efficiency. The model concentrations are considered as related to homogeneous spheres of radius  $r_k$. Thus, Mie's extinction efficiency $Q_\mathrm{ext}$ comes into play:
\be\label{sigma_ext}
\sigma^{(s)}_\mathrm{ext}(r_k ) = \pi r_k^2 \, Q_\mathrm{ext}(r_k; \lambda=550 \mathrm{nm}; m^{(s)})
\ee
Here the dependence on the complex refractive index $m^{(s)} = n^{(s)} -  i\kappa^{(s)}$ of specie $s$ is highlighted and the in-air light wavelength $\lambda$ is kept fixed. The extinction efficiency plays a crucial role in the calculation of AOD since, for $\Re(m)<2$, its value varies from exactly zero to about 6.

\subsection{Effect of moisture}\label{sec:opt_module_hygro_eff}
Some of the atmospheric aerosol species are hydrophilic: above a certain RH they are no more in the solid state but rather form an aqueous solution. Depending on chemical composition, there is either a continuous or an abrupt (deliquescence) transition to the aqueous solution \cite{SP98}. As a result of water uptake, both particle density and refractive index are driven towards liquid water density and refractive index respectively. This  affects particle optical properties twofold.
On the one hand, the aerosol size distribution is shifted towards larger radii as a result of hygroscopic growth. 
On the other hand, the extinction efficiency curve as a function of radius $Q_\mathrm{ext}(r)$ is modified. The position of its maximum is shifted towards a larger value because  wet refractive index is generally lower than dry one.
Which of either effects prevails, extinction enhancing or quenching, depends on the actual size of  peak shift and hygroscopic growth. 

TCAM dry particle radii $r_{k0}$ are rescaled by a  RH-dependent factor, 
\be\label{hygro_growth}
r_k = r_{k0} \cdot \beta^{(s)}(RH)
\ee
Such wet radius $r_k$ is then employed in \Eq{sigma_ext} for the computation of the extinction cross section. At this place it is worth noting that the aerosol number concentrations  $N^{(s)}_k$ are obtained from dry radius $r_{k0}$: 
\be
N^{(s)}_k = M_k^{(s)}  \left( \rho^{(s)}\frac{4}{3}\pi r_{k0}^3 \right)^{-1}
\ee
with the mass densities $\rho^{(s)}$ as in \Tab{tab:aerclasses_LUT}.  $N^{(s)}_k$ are defined on a 3-dimensional grid given by the cartesian product of a 2-dimensional horizontal grid with $n_H$=11 vertical layers of TCAM.

$\beta^{(s)}(RH)$ is taken to be a fit of Tang's calculations \cite{tang96}, which are based on the single particle levitation technique measurements. In doing so, any hysteresis in the growth-evaporation cycle of hydrophilic particles is neglected. A single growth curved has been used also by other authors \cite{hodz04, koel06}.

 As a second step for accounting for moisture, wet complex refractive index $m=n - i\kappa$ is computed from a weighted average of dry aerosol index $m_0$ and liquid water index $m_\mathrm{w}$, weights being the volume fractions in the mixture, 
 \bea\label{wet_refr_idx}
      m (RH) &=& \frac{m_0 P_0 + m_\mathrm{w} P_\mathrm{w}}{P_0 + P_\mathrm{w}} \\ \nonumber
      P_0 &=& 1 \\ \nonumber
      P_\mathrm{w} &=&  [\beta^{(s)}(RH)]^3-1      
 \eea
 (The weights $P_{0,\mathrm{w}}$ are expressed in units of dry particle volume). 
 Volume weighting is chosen, being the simplest of three weighting procedures which are nearly equivalent when applied to the  computation of refractive index \cite{lesi02}.

  A fair fit of the growth factor displayed in \cite{tang96} for a group of inorganic compounds is the following function:
\be\label{growth_factor}
     \beta^{(s)}(RH) = 1 + \gamma^{(s)}\left[\frac{1}{2} RH + \frac{5}{2} (RH)^{12}\right]
\ee
\Eq{growth_factor} is used for RH$\le100\%$. In case of supersaturation,  capping of  hygroscopic effects is applied, i.e. $\beta^{(s)}(RH>100\%)= 1+3\gamma^{(s)}$.
The sole adjustable parameter in \Eq{growth_factor} is $\gamma^{(s)}$, whose values are listed in  \Tab{tab:aerclasses_LUT}. 

% OC:
The hygroscopic properties of organic particles emitted during laboratory
biomass burning experiments were studied in \cite{mark09}. The resulting growth factor strongly depends on the investigated specie. However, TCAM organic aerosols include  alkanes, aromatic compounds, olefins, and cresol. None of them is hydrophilic, since they are not able to bind water or (case of cresol) since they  are not well soluble. Thus, $\gamma^{(s)}=0$ for such species.
\begin{table}
\begin{center}
\begin{tabular}{ccccc|c}\hline
		compound & $\rho$ [g/cm$^3$ ] & n & $\kappa$ & Ref. & $\gamma$ \\
		\hline\hline
		EC & 1.80 & 1.95 & .790 &   \cite{bond06} & 0 \\
		OC & 1.50 & 1.53 & .006 &    \cite{tsig08} & 0  \\
		sulfates & 1.60 & 1.53 & .000 &   \cite{deme07} & 1 \\
		nitrates& 1.60 & 1.53 & .000 &   \cite{deme07} & .75 \\
		ammonium & 1.60 & 1.53 & .000 &    \cite{deme07} & .75 \\
		dust & 2.65 & 1.50 & .033 &    \cite{redd05} & 0 \\				
		\hline					
\end{tabular}	
\caption{{\it Mass density $\rho$,  real $n$ and imaginary $\kappa$ part of dry refractive index (at wavelength $\lambda$=550 nm), hygroscopicity parameter $\gamma$ (see \Eq{growth_factor}), for each of the aerosol compounds considered in this work.}}
\label{tab:aerclasses_LUT}
\end{center}
\end{table}

\subsection{Cloud screening}\label{sec:opt_module_cloudmask}
MODIS retrieval are not allowed over cloudy or foggy pixels by the AOD algorithm \cite{levy07}. Conversely, model AOD is available even in case of total cloud coverage. Thus, for making satellite and model AOD more directly comparable, a cloud mask is applied to model AOD too. The diagnostic variable Cloud Water Content (CWC) is employed, which is obtained from MM5 outputs. The screening protocol consists in requiring null CWC over the whole vertical profile.  The protocol ensures that all cloud contaminated pixels are screened out from the model.

\section{The case study}\label{sec:results}
The method outlined in \Sect{sec:opt_module} has been applied to TCAM aerosol simulations from January 1 to December 31, 2004. 
The study area covers nearly whole  Northern Italy, which is divided into 41x64 cells (10x10 km$^2$ resolution) In the following this area is also termed "QUITSAT domain".

Simulations have been performed getting initial and boundary conditions by a nesting procedure from the results of CHIMERE simulations over Europe \cite{schm01}. The emission fields were estimated by POEM-PM pre-processor starting from the regional emission inventory collected by the ISPRA (Istituto Superiore per la  Protezione e la Ricerca Ambientale).  The point sources and area emissions have been chemically and temporally adjusted to TCAM requirements and have been derived by CTN-ACE Italian modeling intercomparison project 
%\cite{ctn-ace}
. Meteorological fields have been created with MM5 model and adapted to TCAM through PROMETEO pre-processor.

Hourly aerosol concentrations from TCAM QUITSAT v.3 simulations have been employed as an input to the AOD computation. TCAM delivers the concentrations of 21 species, which have been aggregated into $n_S=6$ chemical compounds. One of these is unbound or elemental carbon,  "EC". Primary and secondary organic carbon species have been lumped into the "OC" class. Then,  some inorganic electrolytes have been neglected for the sake of lighten AOD computation, since they exhibit very low concentrations in TCAM QUITSAT simulations: They include sea-salt ionic components Na$^+$ and Cl$^-$ (cp. \cite{deme07}),  aqueous phase SO$_2$(aq), H$_2$O$_2$(aq), O$_3$(aq), and  protons H$^+$. Liquid water has not been considered as an aerosol specie for its own. However, the role of water is accounted for through the parametrization of hygroscopic growth according to \Eq{growth_factor}. Finally, one is left with the 6 aerosol compounds, as listed in \Tab{tab:aerclasses_LUT}. 

\subsection{Size classes}\label{sec:results_sizeclasses}
\bfi
\centering
\begin{minipage}{.5\linewidth}
\includegraphics*[width=7.cm] {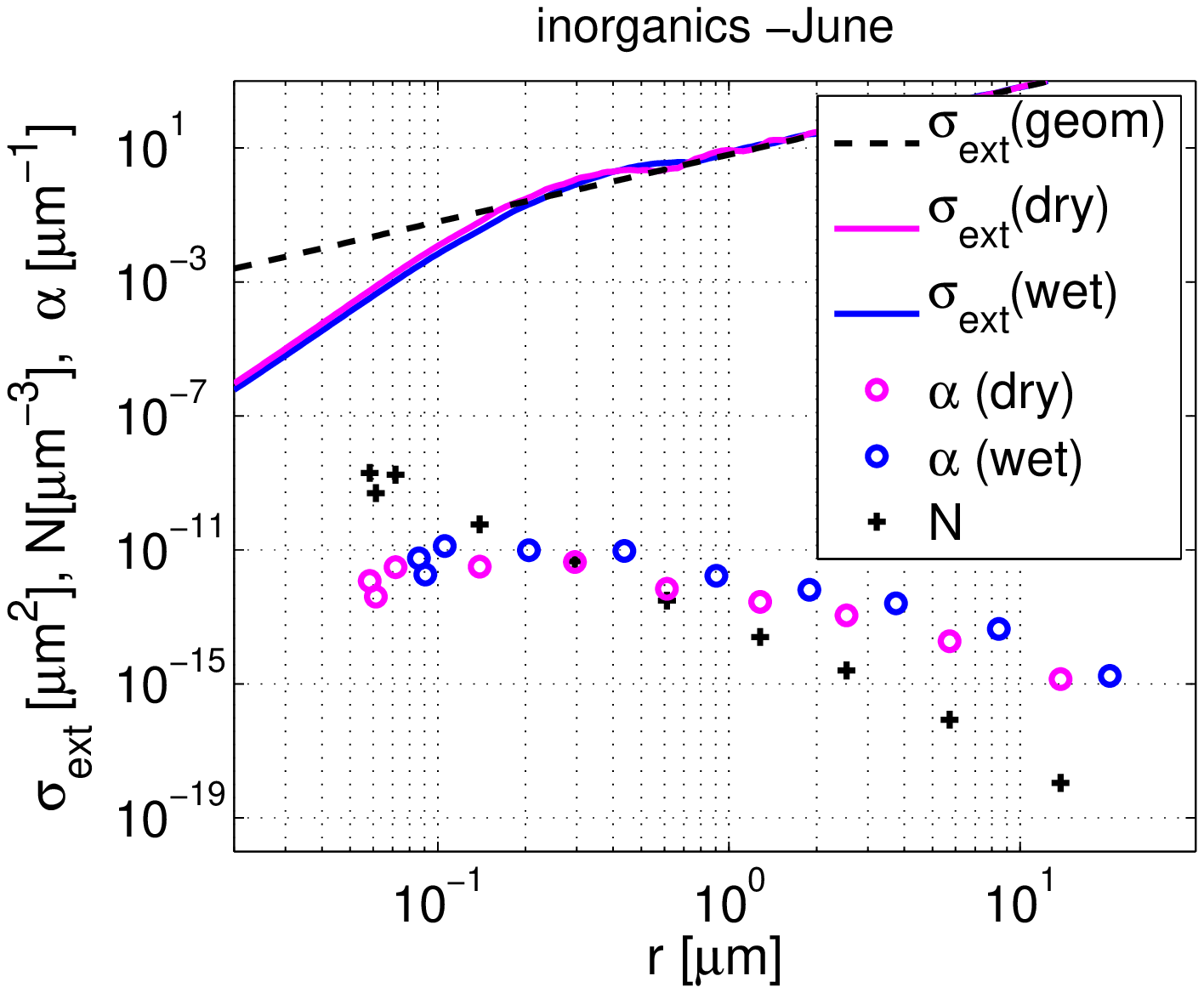}
\end{minipage}\hfill
\begin{minipage}{.5\linewidth}
\includegraphics*[width=7.cm] {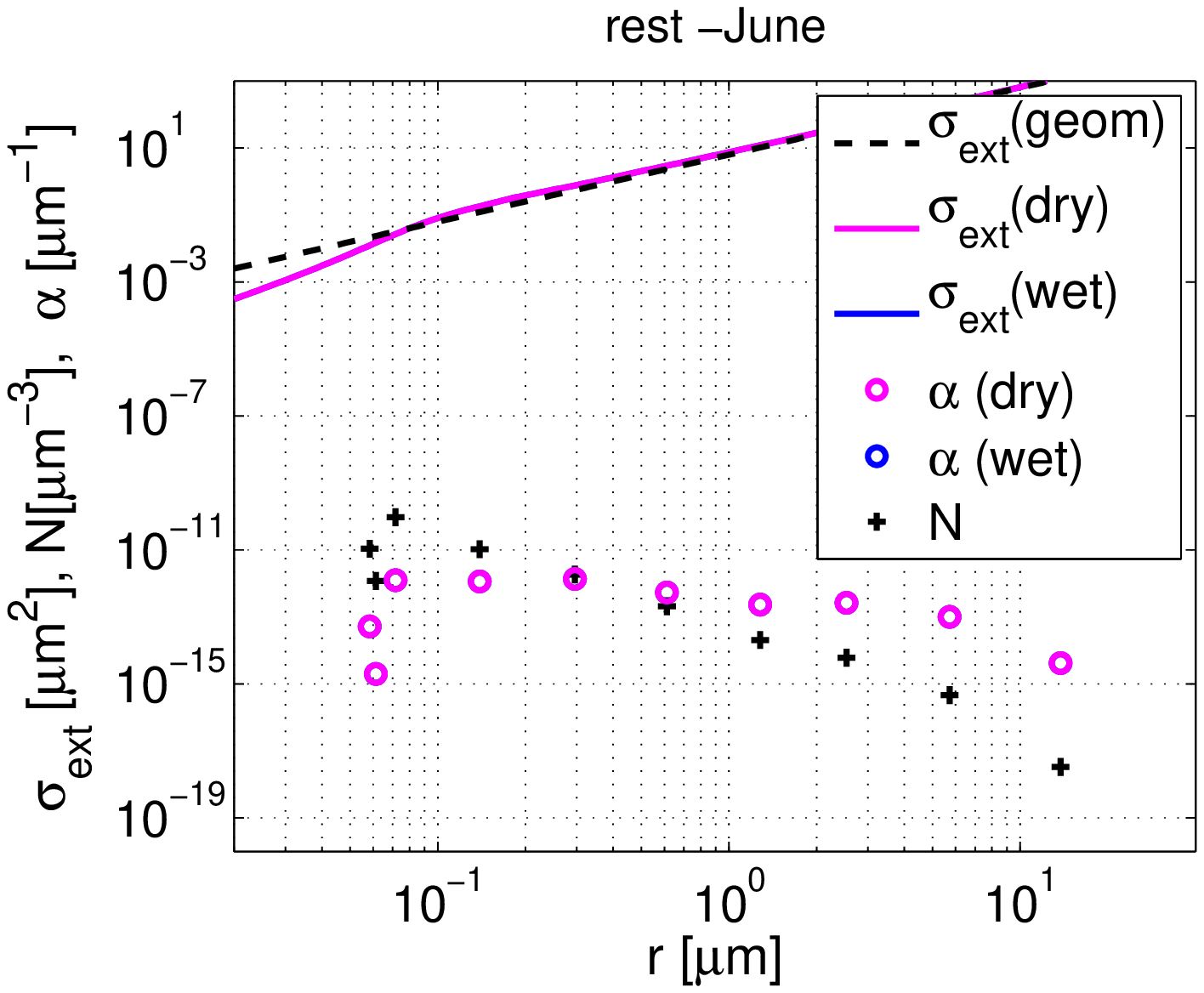}
\end{minipage}
\begin{minipage}{.5\linewidth}
\includegraphics*[width=7.cm] {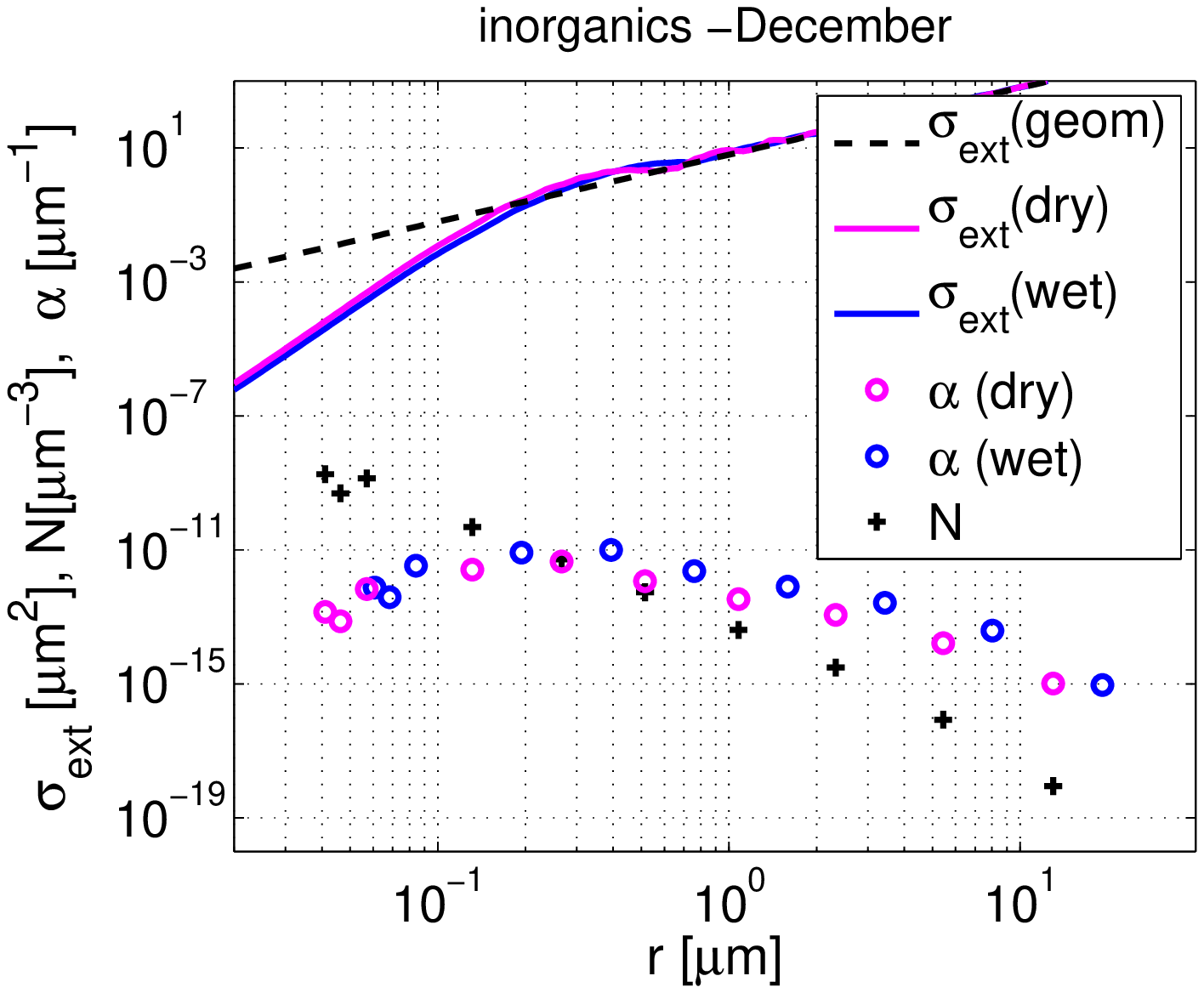}
\end{minipage}\hfill
\begin{minipage}{.5\linewidth}
\includegraphics*[width=7.cm] {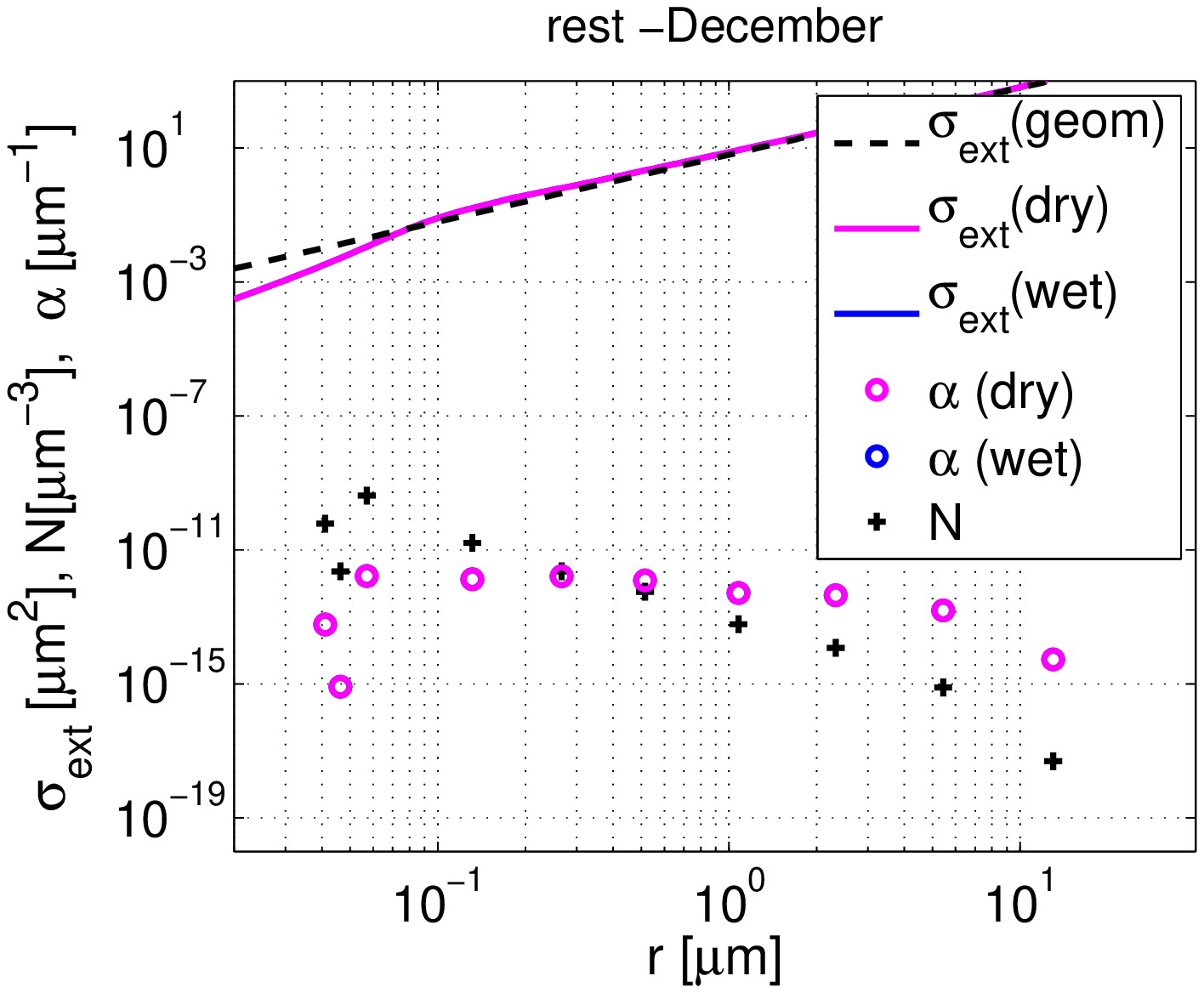}
\end{minipage}
\caption{{\it TCAM number concentrations $N_K$ (crosses); extinction cross section $\sigma_\mathrm{ext}(r_k)$ and its geometrical limit (lines); extinction coefficient components  $\alpha_k$ (circles) for both dry (magenta) and wet (blue) case. Wet case is obtained considering RH=80\%. Left panels: inorganic compounds (ammonium, nitrates, sulfates); Right panels: remaining compounds (EC, OC, dust). The remaining compounds are not hygroscopic, thus dry and wet curves overlay. For this group, the EC extinction cross section is displayed. Values in top (bottom) panels refer to the  3-dimensional domain average for the month  June(December) 2004 considering times between 9:00 and 14:00 UTC.}}  
\label {fig:NCalpha}
\efi
Each of the 6 chemical compounds is organized into $n_K$=10 dimensional dimensional classes, whose size is different for each simulation pixel and for each time step, due to the specific fixed-moving algorithm for TCAM size distribution. In \Fig {fig:NCalpha}  domain mean size distributions averaged over time spans of one month are displayed. In these plots, the 6 chemical compounds are further lumped into an inorganic group (ammonium, nitrates, sulfates) and the remaining compounds (EC, OC, dust).  Summer and winter typical situations are displayed.  The 4 smallest dimensional classes (radii $<$ 140nm) are always the most populated in terms of  number concentration $N_k$. The size  distribution of the inorganic group nearly follows a power-law distribution, while the distribution of the remaining compounds is peaked at a radius below 100 nm. This holds both in June and December, 2004. It is also seen that the smallest 3 classes are affected by the most significant changes between June and December, while the larger classes are rather insensitive to the choice of the month. The general trend is that larger radii are obtained in summer, independently of the chemical group. This behavior should be related to photo-induced aging of primary pollutants.

Mie's extinction cross section $\sigma_\mathrm{ext}(r, \lambda=550$nm$, m)$  with $m=1.53-i0 \, (1.95-i0.79)$ for the inorganic(rest) group is also displayed in \Fig {fig:NCalpha}. It is obtained from a code by C. M\"atzler (Universit\"at Bern), publicly available on the web (www.iap.unibe.ch/publications/download/201/en/).  The asymptotic behaviors of Mie's efficiency are exploited for lighten numerical computations. A look-up table of  $Q_\mathrm{ext}(r, n, \kappa)$ for logarithmically spaced radii $r$ and imaginary parts  $\kappa$ and for linearly spaced real parts of refractive index $n$ is linearly interpolated at each model cell, at each size bin, at each time step. A look-up table approach to $Q_\mathrm{ext}$ is employed in the optical module of the WRF-chem model too \cite{fast05}.

If the cross section $\sigma_\mathrm{ext}(r_k)$ gets multiplied by the aerosol number concentration $N_k$, the  class-resolved extinction coefficient $\alpha_k$ is obtained. Since  $\sigma_\mathrm{ext}(r_k)$  is (nearly) logarithmically increasing and $N_k$ is  logarithmically decreasing for particle class-radius $r_k>80$nm, the product of both has got a maximum value at intermediate values. Thus, the most significant contributions to the total extinction coefficient result  from particles whose radius is below 0.5$\mu$m. Larger values of the extinction are obtained for the inorganic group, which is a consequence of larger population of this kind of aerosols according to TCAM. 

In \Sect{sec:opt_module_hygro_eff}  it has been mentioned that, in general, moisture may lead  to both  extinction  enhancing or quenching.  In \Fig {fig:NCalpha} it is  seen that in the actual case moisture results in enhanced extinction coefficient for every dimensional class. This is due to the fact that most of the aerosol population is submicron sized.

\subsection{Vertical structure}
\bfi
\centering
\begin{minipage}{.5\linewidth}
\includegraphics*[width=7.cm] {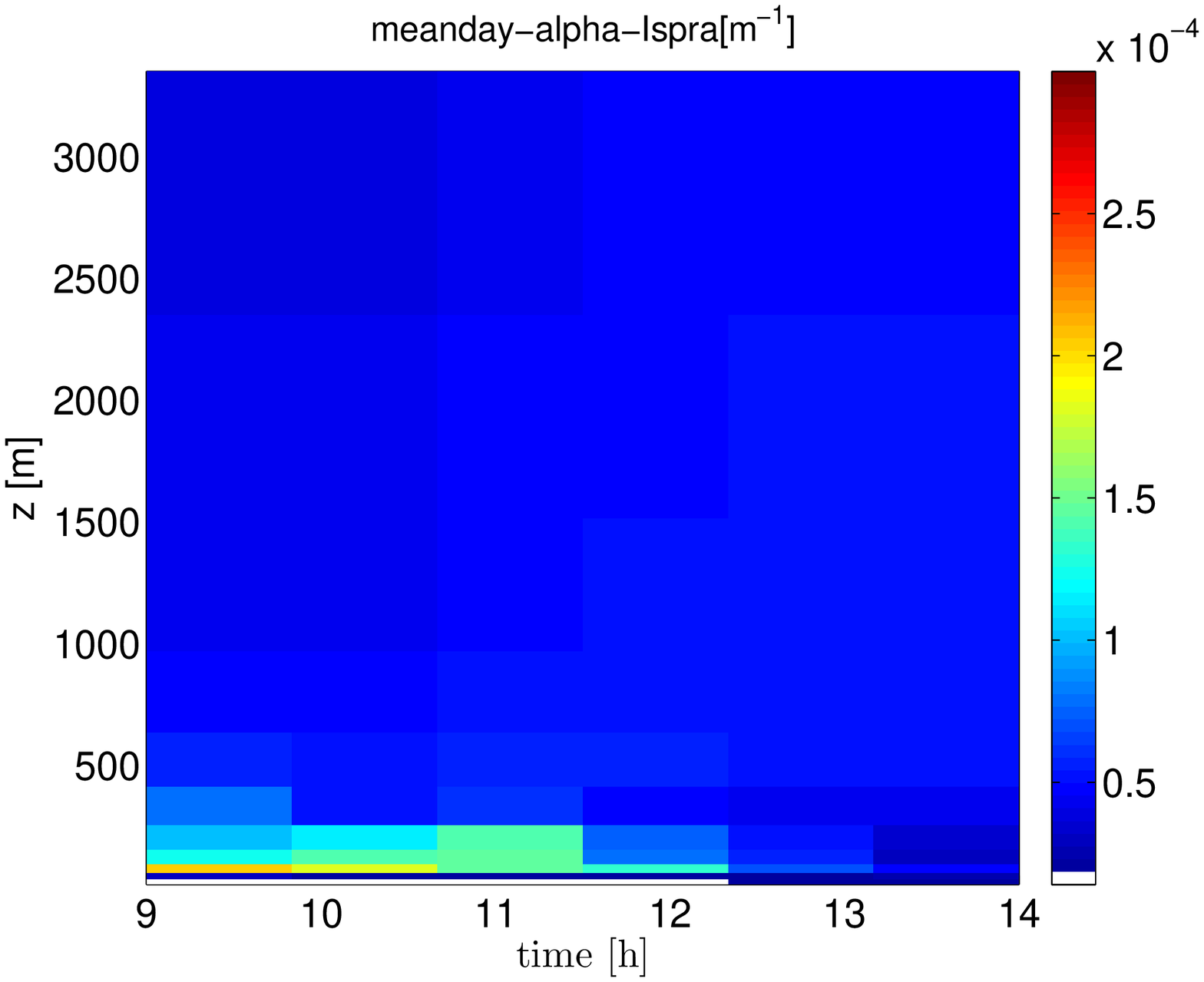}
\end{minipage}\hfill
\begin{minipage}{.5\linewidth}
\includegraphics*[width=7.cm] {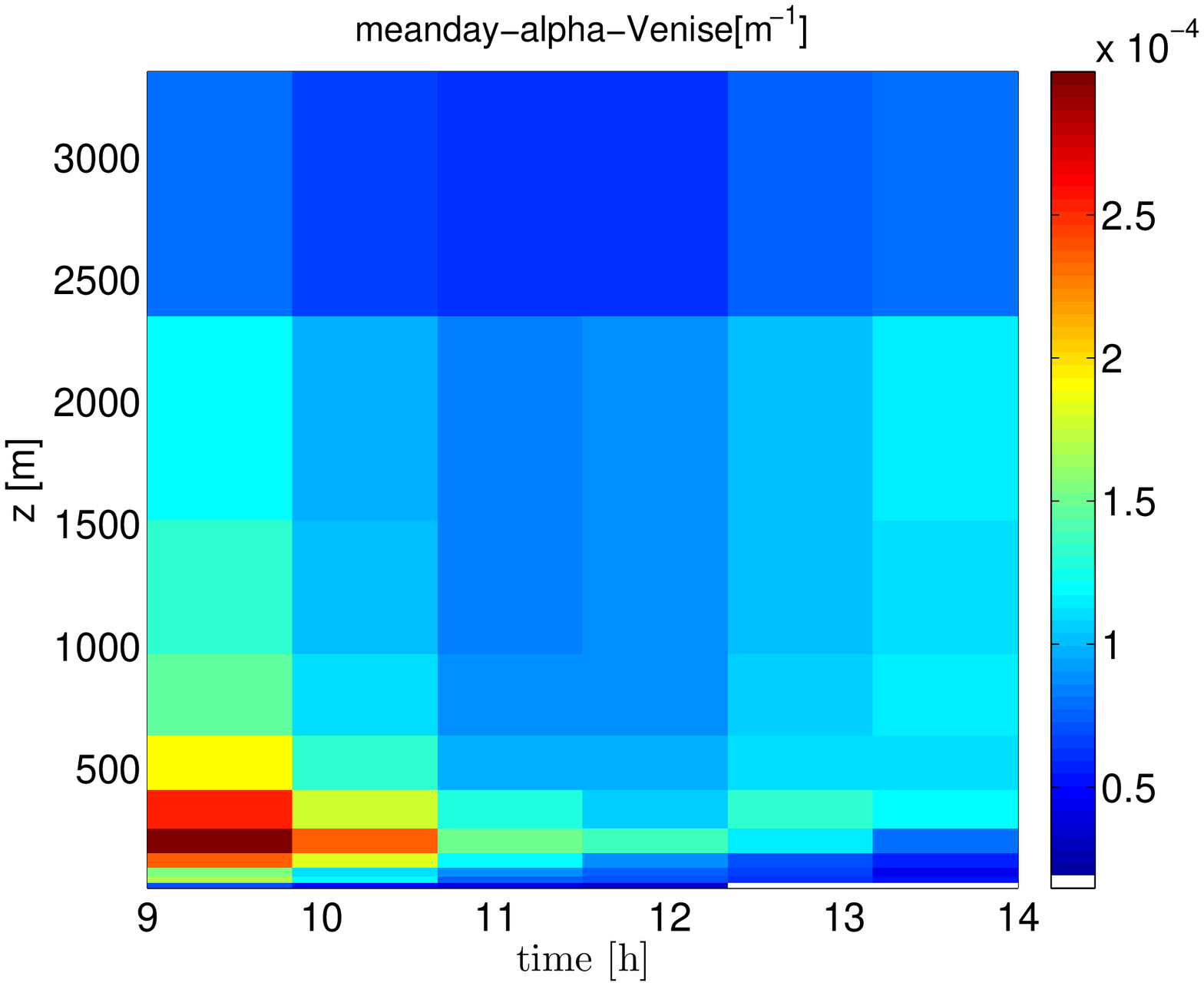}
\end{minipage}
\caption{ {\em Temporal evolution between 9:00 and 14:00 UTC of  vertical profile of extinction coefficient $\alpha(t,z)$ for mean day in June 2004, accounting for moisture. Left(right) panel: model cell next to AERONET site Ispra(Venise).}}\label {fig:alpha_profiles}
\efi
The temporal evolution of the vertical resolved extinction coefficient $\alpha(z)$ is displayed in \Fig {fig:alpha_profiles} for the mean day in June 2004.  Data refer to  the domain cells which are next to Ispra and Venise AERONET sites respectively. Largest $\alpha(z)$ values are obtained at heights $z<500$m. For the highest layers, the extinction is about 1/10 of the maximum value assumed in each vertical profile.  This means that the vertical modeling within TCAM is able to capture most of the optical extinction responsible for AOD (June usually is the month during which the insulation and thus the boundary layer height reach their maximum).  In any case, $\alpha(z=h_{11})$ values for TCAM QUITSAT simulations are quite low and are comparable to the extinction resulting from molecular scattering in a clean atmosphere ($\sim1.3\times10^{-5} {\mathrm m}^{-1}$, \cite{SP98}).

\subsection{Comparisons of spatial patterns}
\bfi
\centering
\begin{minipage}{.5\linewidth}
\includegraphics*[width=7.cm] {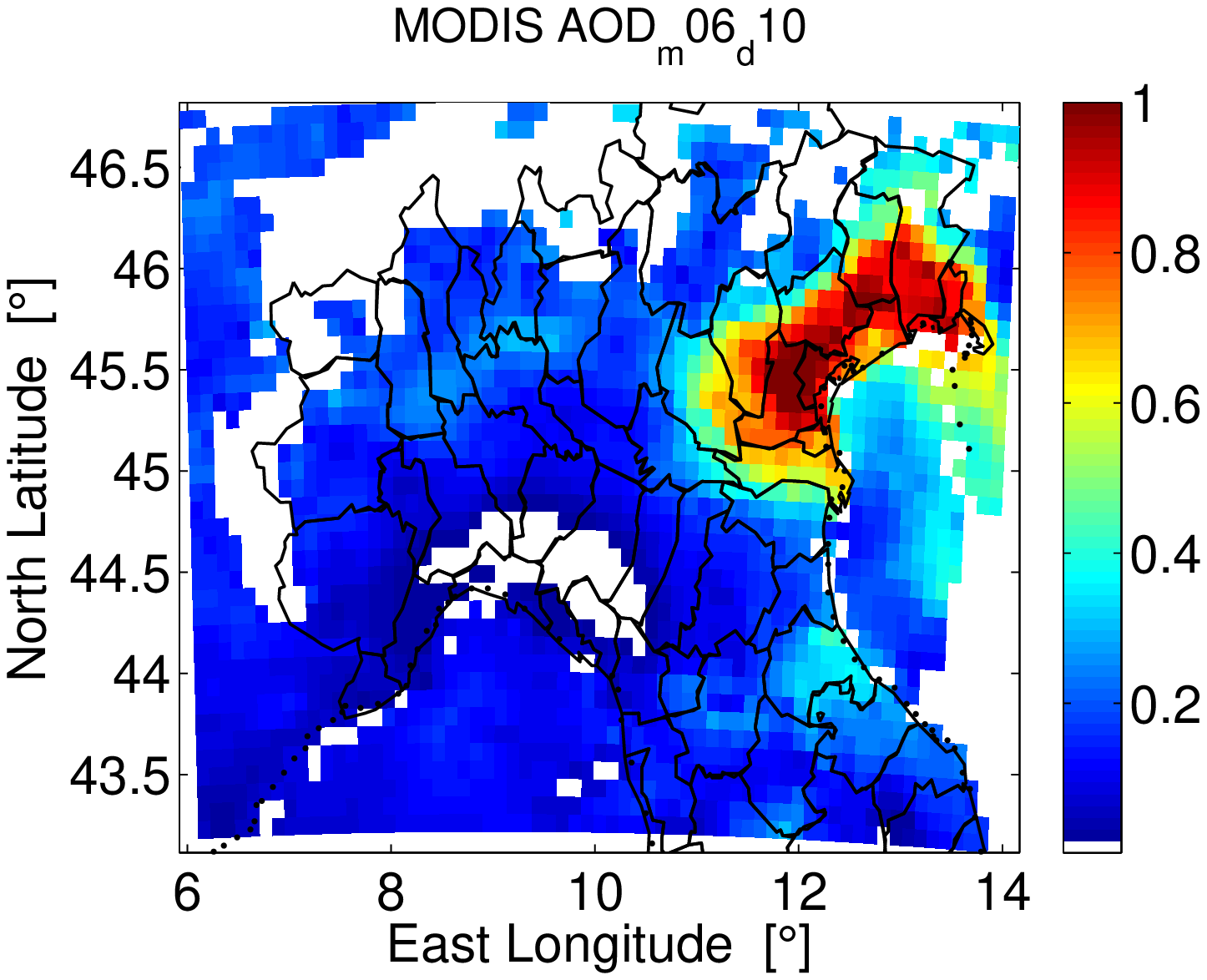}
\end{minipage}\hfill
\begin{minipage}{.5\linewidth}
\includegraphics*[width=7.cm] {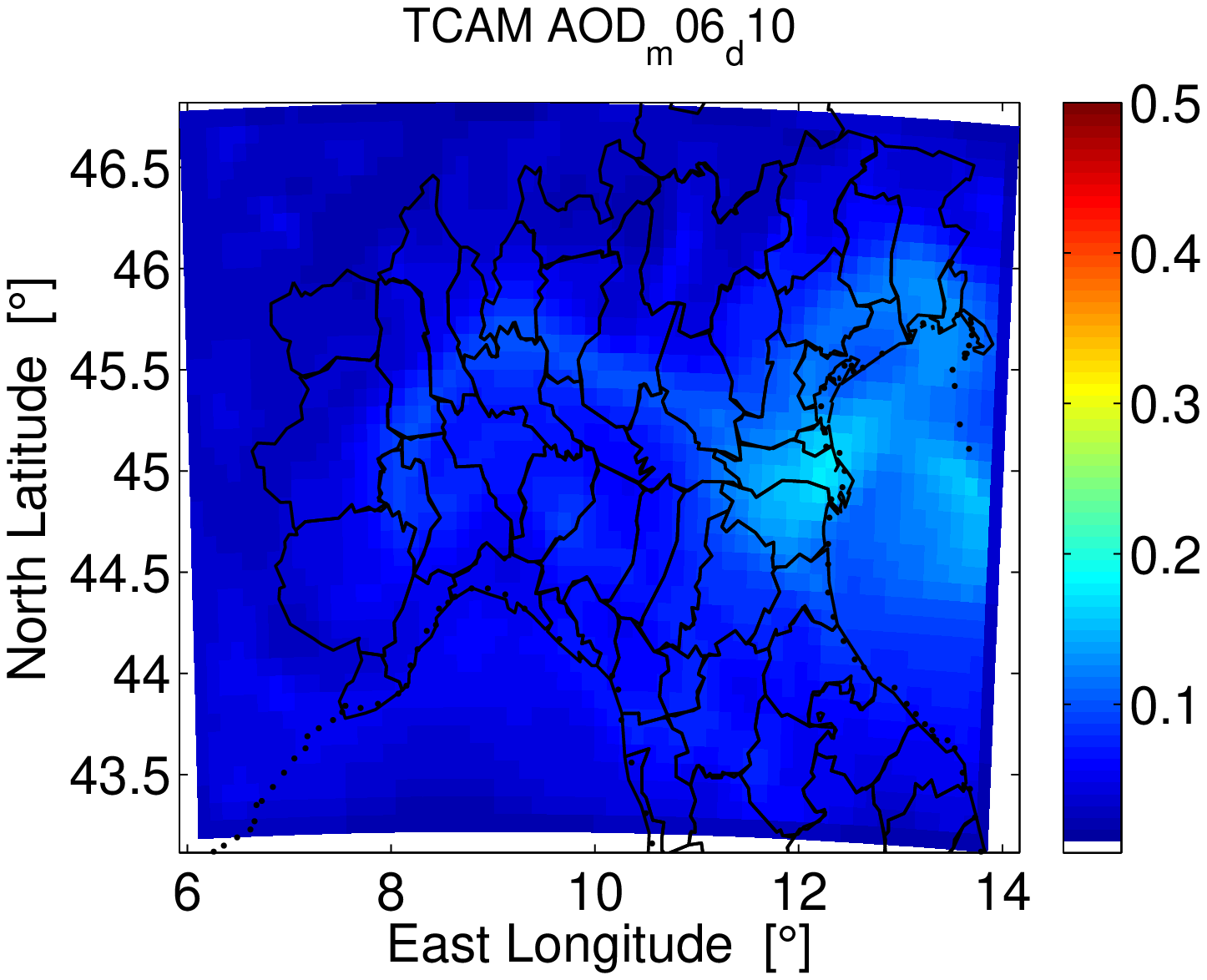}
\end{minipage}
\begin{minipage}{.5\linewidth}
\includegraphics*[width=7.cm] {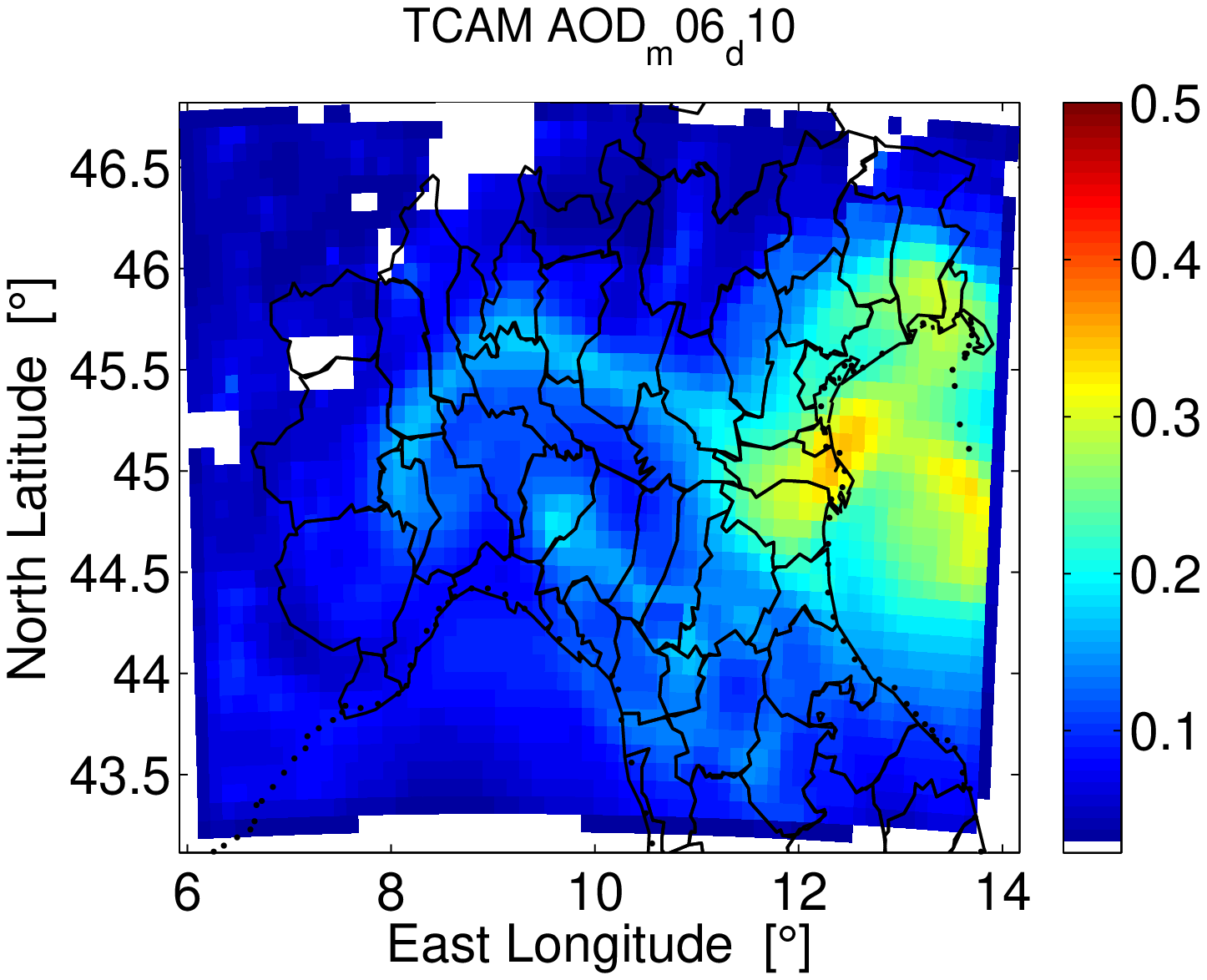}
\end{minipage}\hfill
\begin{minipage}{.5\linewidth}
\includegraphics*[width=7.cm] {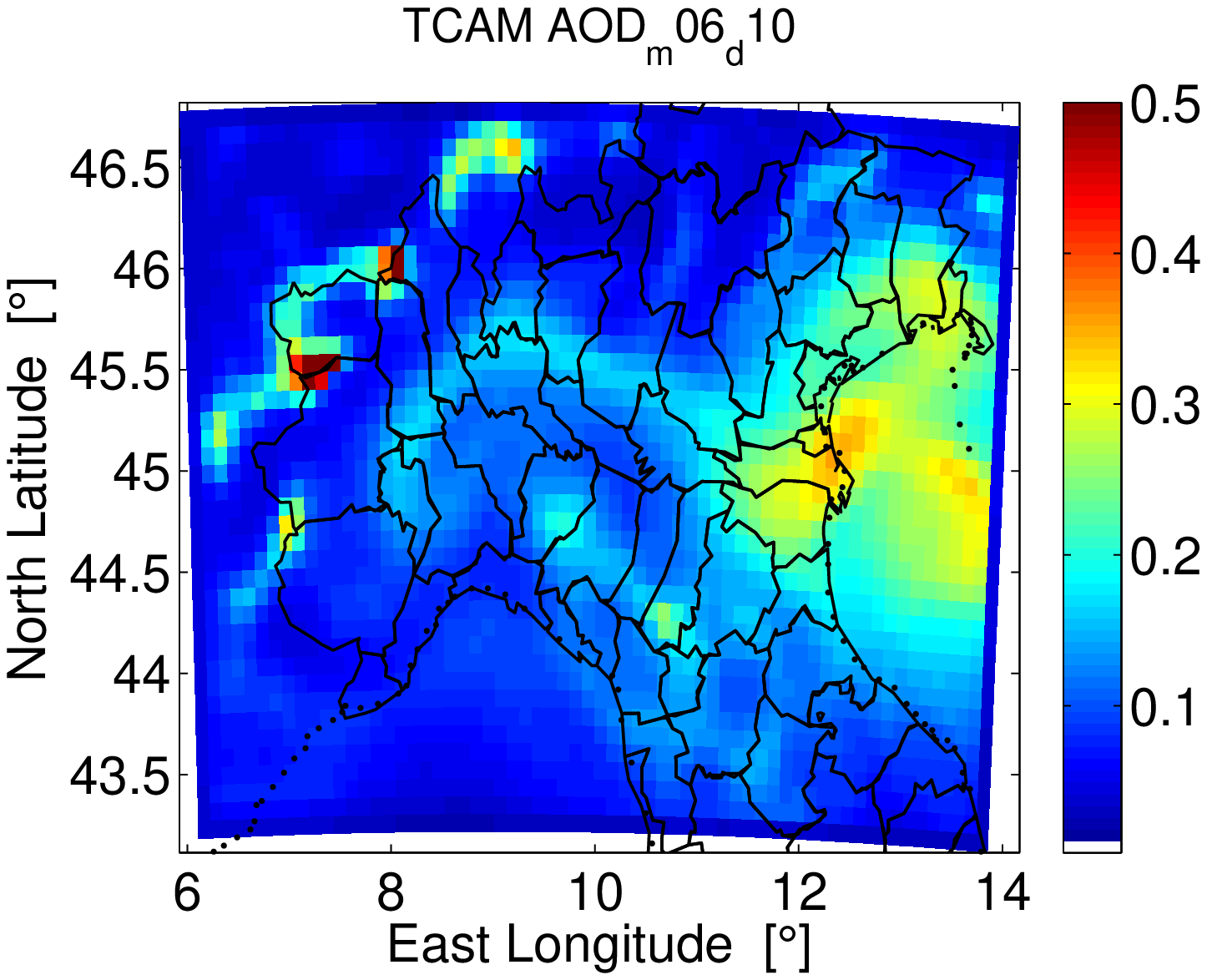}
\end{minipage}
\caption{ {\em AOD maps for June 10, 2004 over the QUITSAT domain. Top-left panel: MODIS product ``Corrected\_Optical\_Depth\_Land\_And\_Ocean'' from both Terra and Aqua overpasses; Top-right panel: TCAM AOD averaged over the times corresponding to the satellite overpasses; Bottom-right: TCAM AOD accounting for moisture; Bottom-left: TCAM AOD accounting for moisture and for  mask. Notice the vertical scale in MODIS AOD panels.}}\label {fig:jun10_AOD}
\efi
According to \Eq{AOD_def}, TCAM AOD is obtained from integration of the vertical profiles of the extinction coefficient.  Additionally, AOD can be obtained from satellite observations too.  MODIS product "Corrected\_Optical\_Depth\_Land" is considered in the following, which is the quality reference for quantitative studies \cite{levy07}. This product is available at the wavelength of 550 nm. Data from both Terra and Aqua satellite are considered. Daily mean AOD maps are obtained upon averaging over all scenes having finite spatial overlap with the QUITSAT domain.

In the first panel of \Fig {fig:jun10_AOD}, such a map is shown for the day June 10, 2004. This is a cloud-free day for most of the domain pixels, with exception of the alpine region and the western part of  Tusco-Emilian Apennine. A strong aerosol signal arises from the eastern part of the domain, with AOD values close to 1 over Veneto and Friuli. Also, AOD $\approx$ 0.4 is observed southern of the western Alps, which is nearly double as the value in the rest of western Po-Valley.
Corresponding model AOD can be obtained averaging over all model maps whose simulation time corresponds within 30 minutes to the time of satellite overpasses.  The results of this procedure are shown in the top-right panel of \Fig {fig:jun10_AOD}. In contrast to satellite map,  model map is defined for each domain cell. Its main feature is a hot spot extending from southern part of Veneto to eastern part of Emilia-Romagna. Then,  an AOD arc southern of western Alps is present. AOD values are a factor of 2-5 lower than in the satellite map. However, the above described qualitative features of the map are present both in model and observations. 

The model output considered so far does not account for moisture. Its effect is computed as in \Sect{sec:opt_module_hygro_eff} and can be appreciated in  TCAM map in the bottom-right panel of \Fig {fig:jun10_AOD}. It is clearly seen that  moisture leads in this case to AOD increase, which is now nearly double  with respect to the dry map. The  importance of moisture for haze over Po-Valley has been already highlighted by \cite{deme07}. Furthermore, quite high AOD values on western Alps and Apennines cells are found. The origin of these high values should be related to Cloud  Optical Depth. Indeed, in the bottom-left panel of \Fig {fig:jun10_AOD}, where cloud masking procedure as in \Sect{sec:opt_module_cloudmask} is applied, many of these pixels are absent.

Finally, it is interesting to compare monthly mean AOD maps from MODIS and TCAM. In \Fig {fig:monthly_mean_AOD} the months June and December 2004 are considered. Within this figure, TCAM maps account for both moisture and cloud mask. In June, while satellite AOD is a factor about 2 larger than model AOD, several spatial patterns are recognized in both maps: the eastern Po-Valley maximum, the Alps and Apennines minima, the local maximum over the Gulf of Genoa, the cloud coverage over western and central Alps. In December, AOD values are lower than in June and the departures between satellite and model are smaller. However, the number of missing values in satellite pictures is much larger than in model computation. This could be due to two reasons: i) snow coverage over Alps; ii) fog in the Po Valley. While model AOD is delivered in both cases, both  meteorological situations lead to failure of the MODIS retrieval algorithm. 
\bfi
\centering
\begin{minipage}{.5\linewidth}
\includegraphics*[width=7.cm] {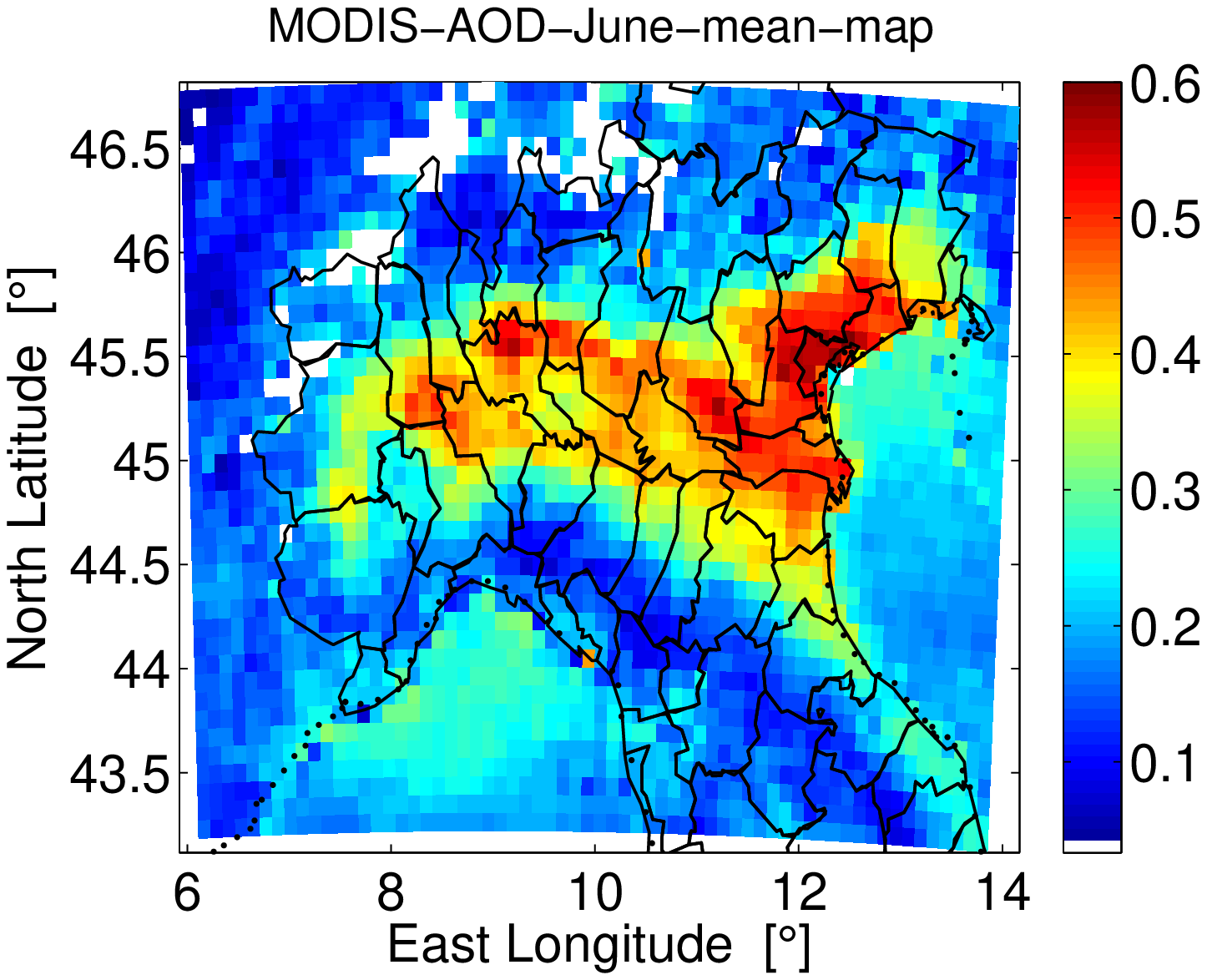}
\end{minipage}\hfill
\begin{minipage}{.5\linewidth}
\includegraphics*[width=7.cm] {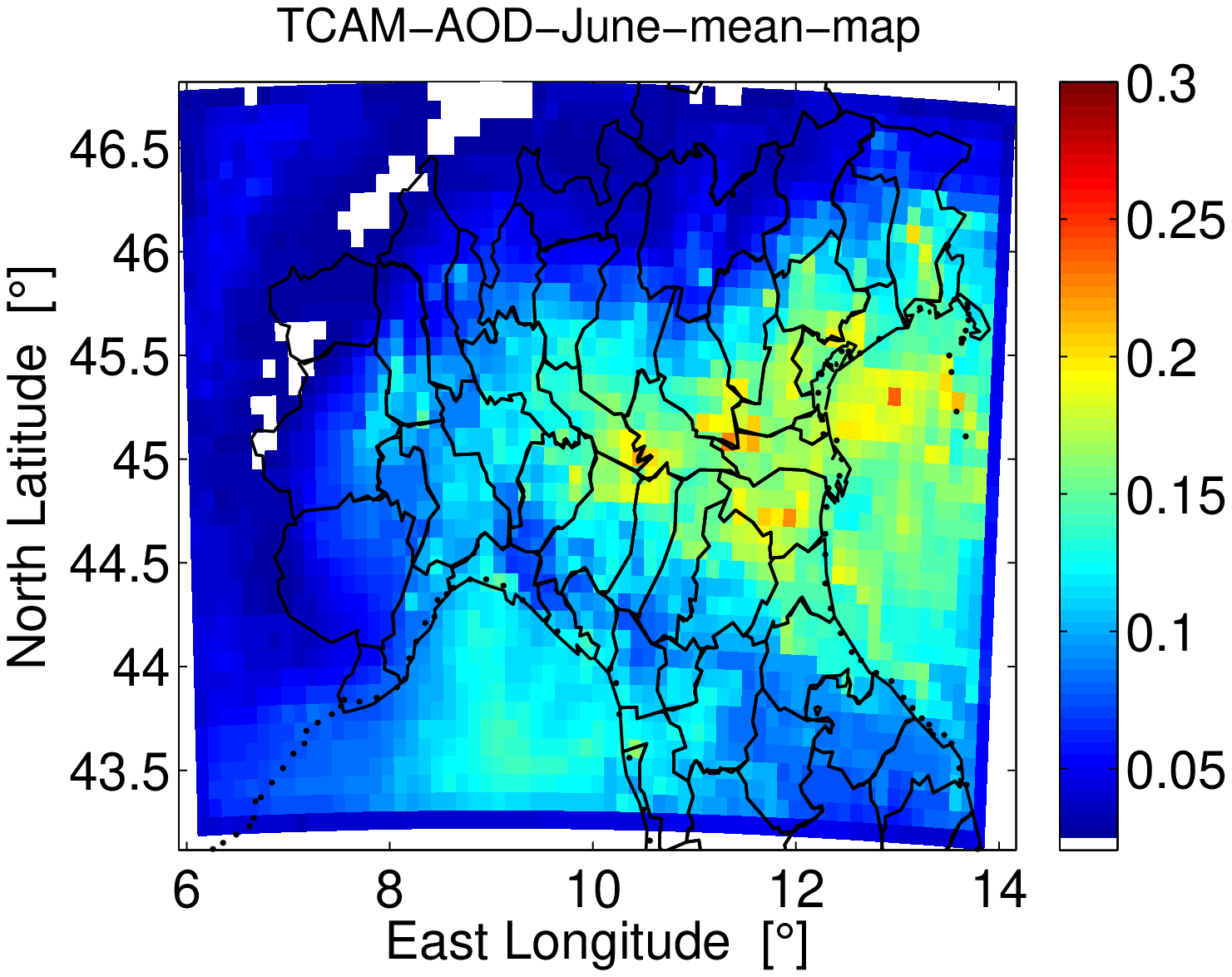}
\end{minipage}
\begin{minipage}{.5\linewidth}
\includegraphics*[width=7.cm] {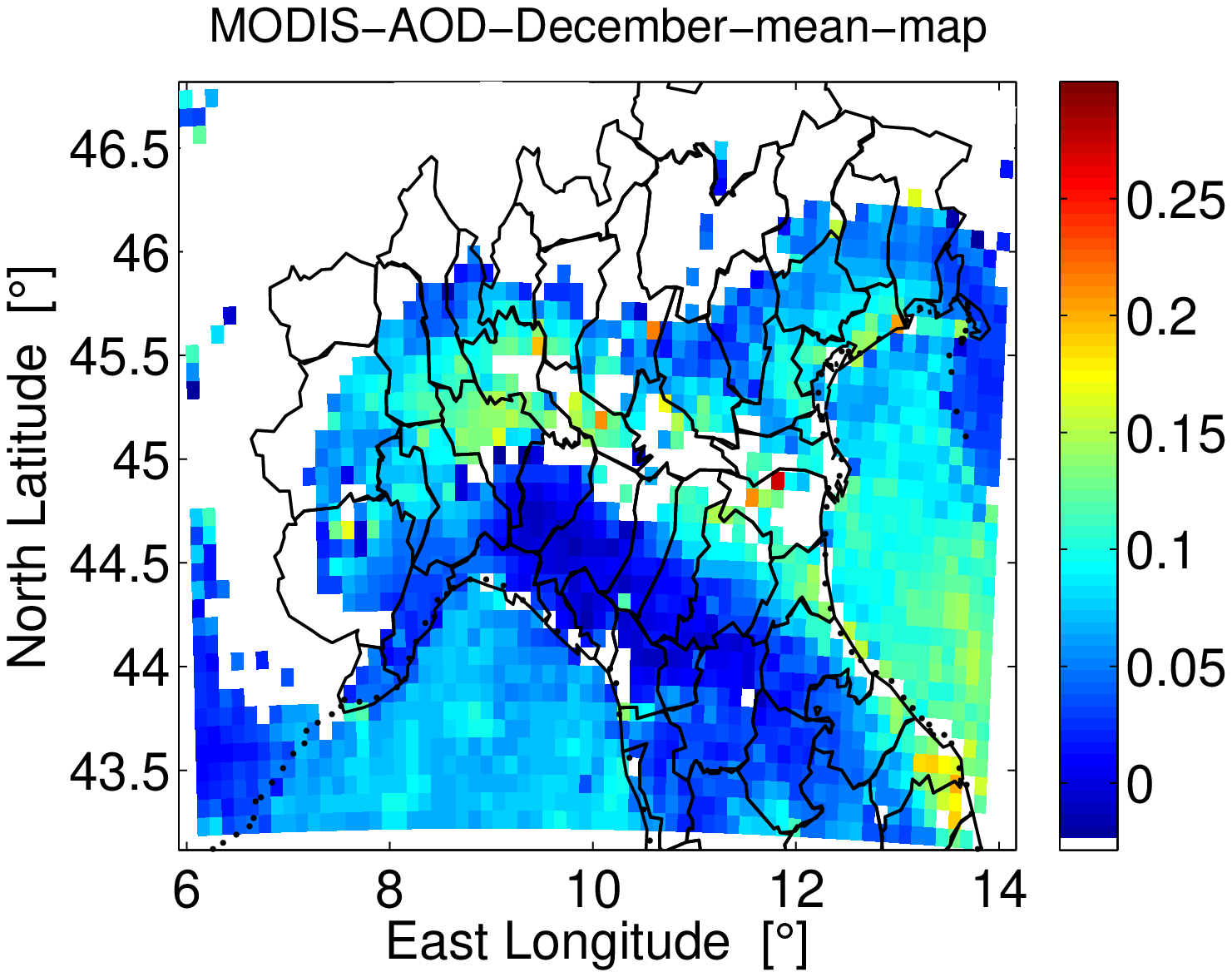}
\end{minipage}\hfill
\begin{minipage}{.5\linewidth}
\includegraphics*[width=7.cm] {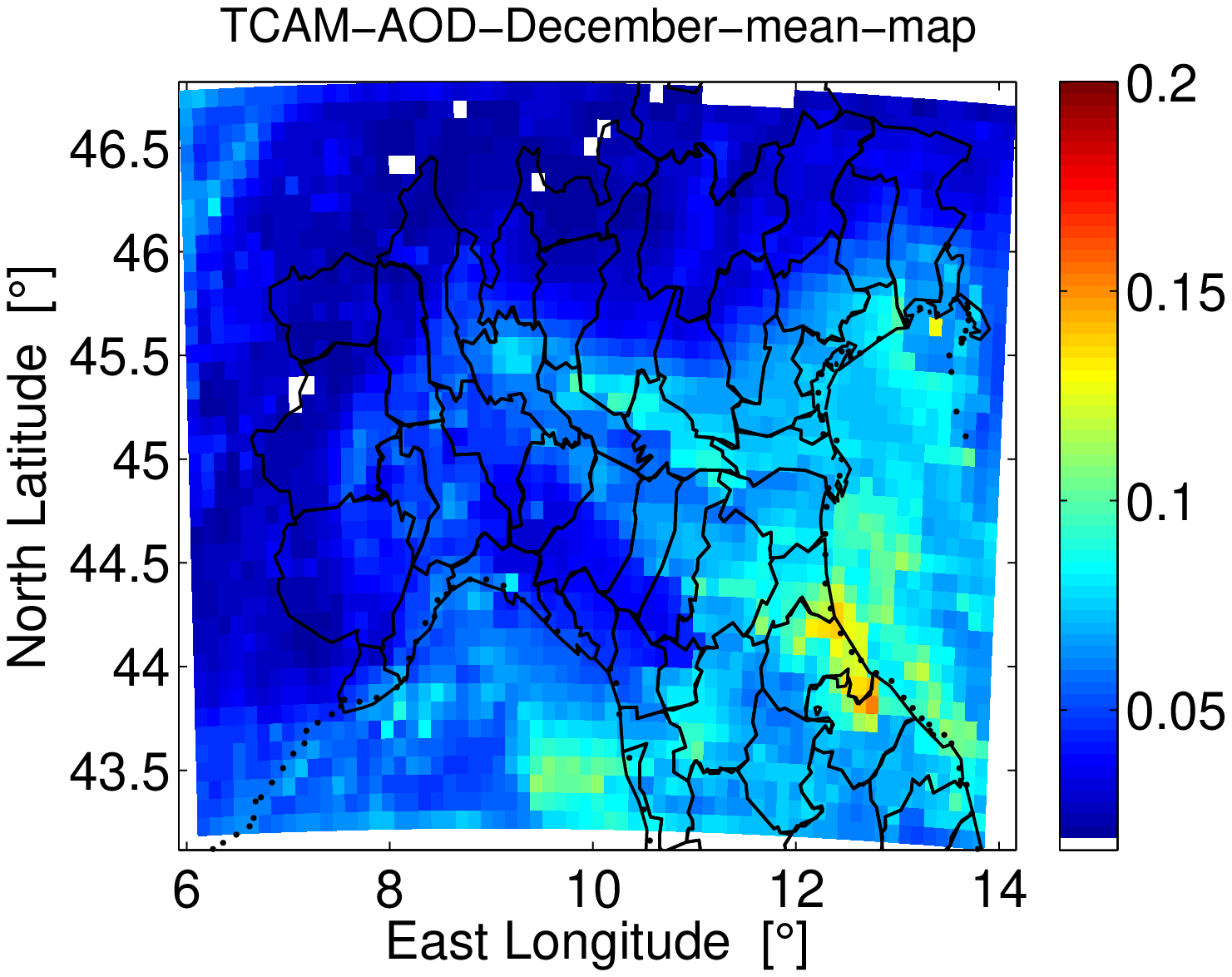}
\end{minipage}
\caption{ {\em AOD mean maps for June (upper row) and December (lower row), 2004 over the QUITSAT domain. Left panels: MODIS product ``Corrected\_Optical\_Depth\_Land\_And\_Ocean'' from both Terra and Aqua overpasses; Right panels: TCAM AOD averaged over the times corresponding to the satellite overpasses accounting for moisture and cloud mask. Notice variable color scales.}}\label {fig:monthly_mean_AOD}
\efi

\subsection{Comparisons of timeseries}
\bfi
\centering
\begin{minipage}{.5\linewidth}
\includegraphics*[width=7.cm] {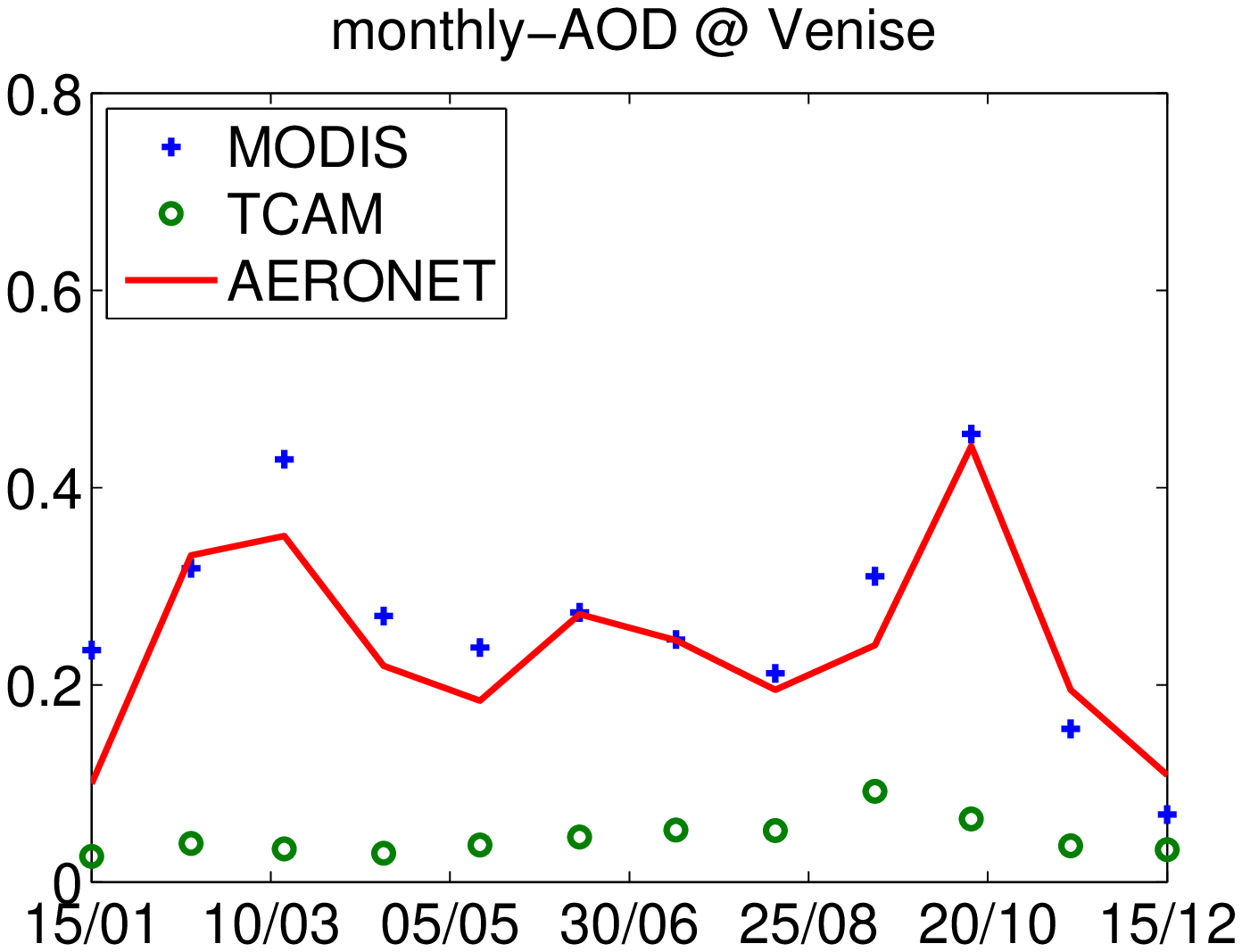}
\end{minipage}\hfill
\begin{minipage}{.5\linewidth}
\includegraphics*[width=7.cm] {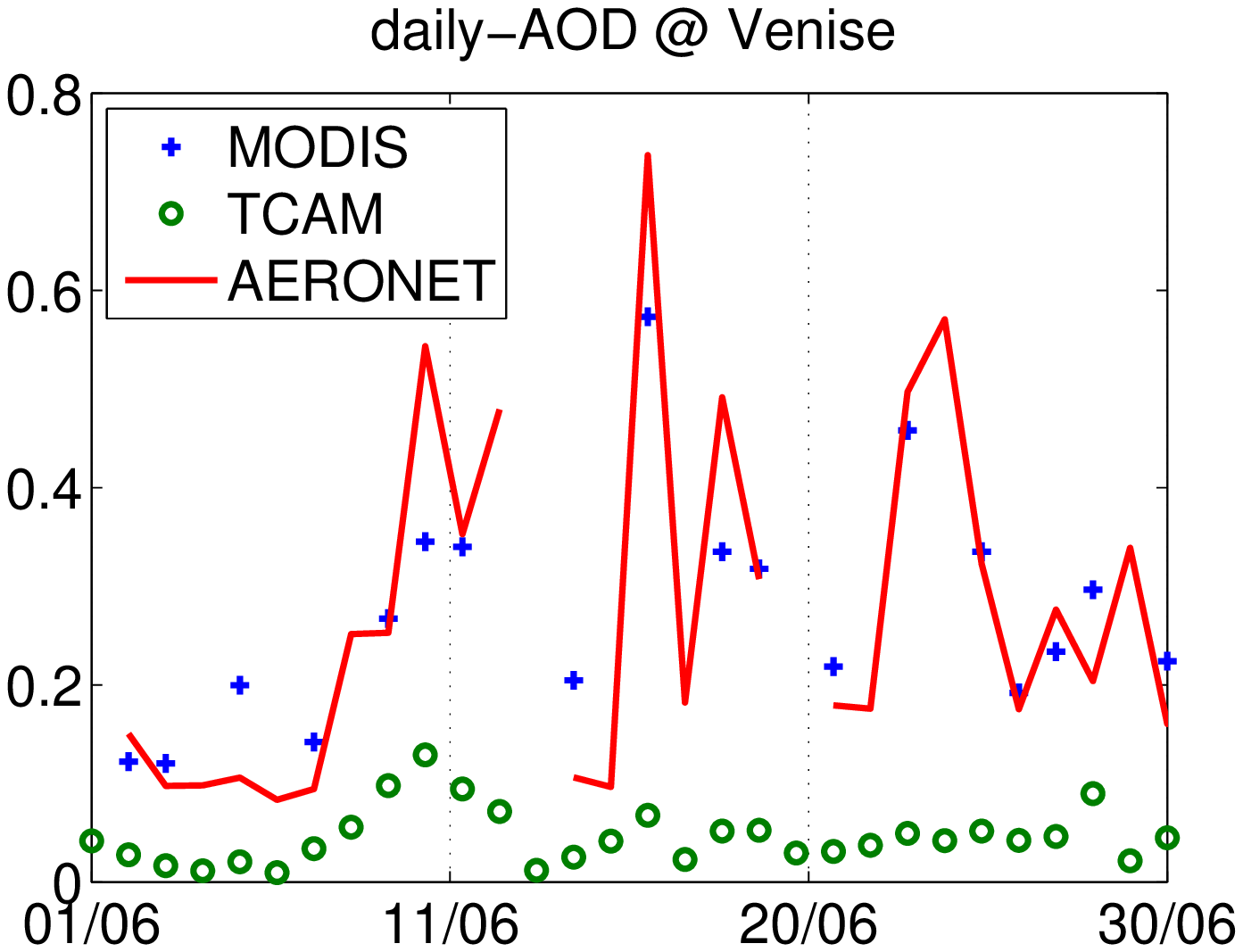}
\end{minipage}\hfill
\begin{minipage}{.5\linewidth}
\includegraphics*[width=7.cm] {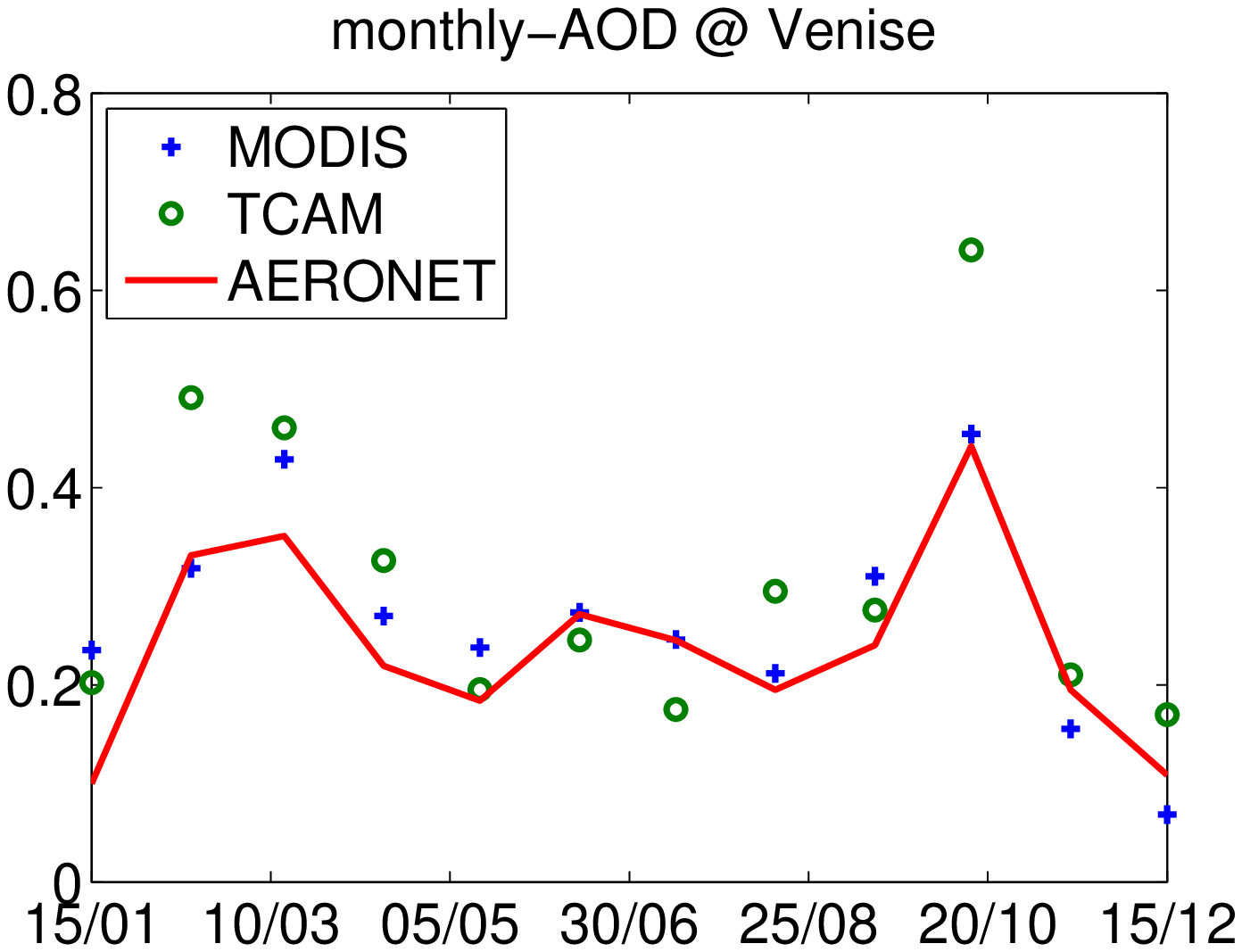}
\end{minipage}\hfill
\begin{minipage}{.5\linewidth}
\includegraphics*[width=7.cm] {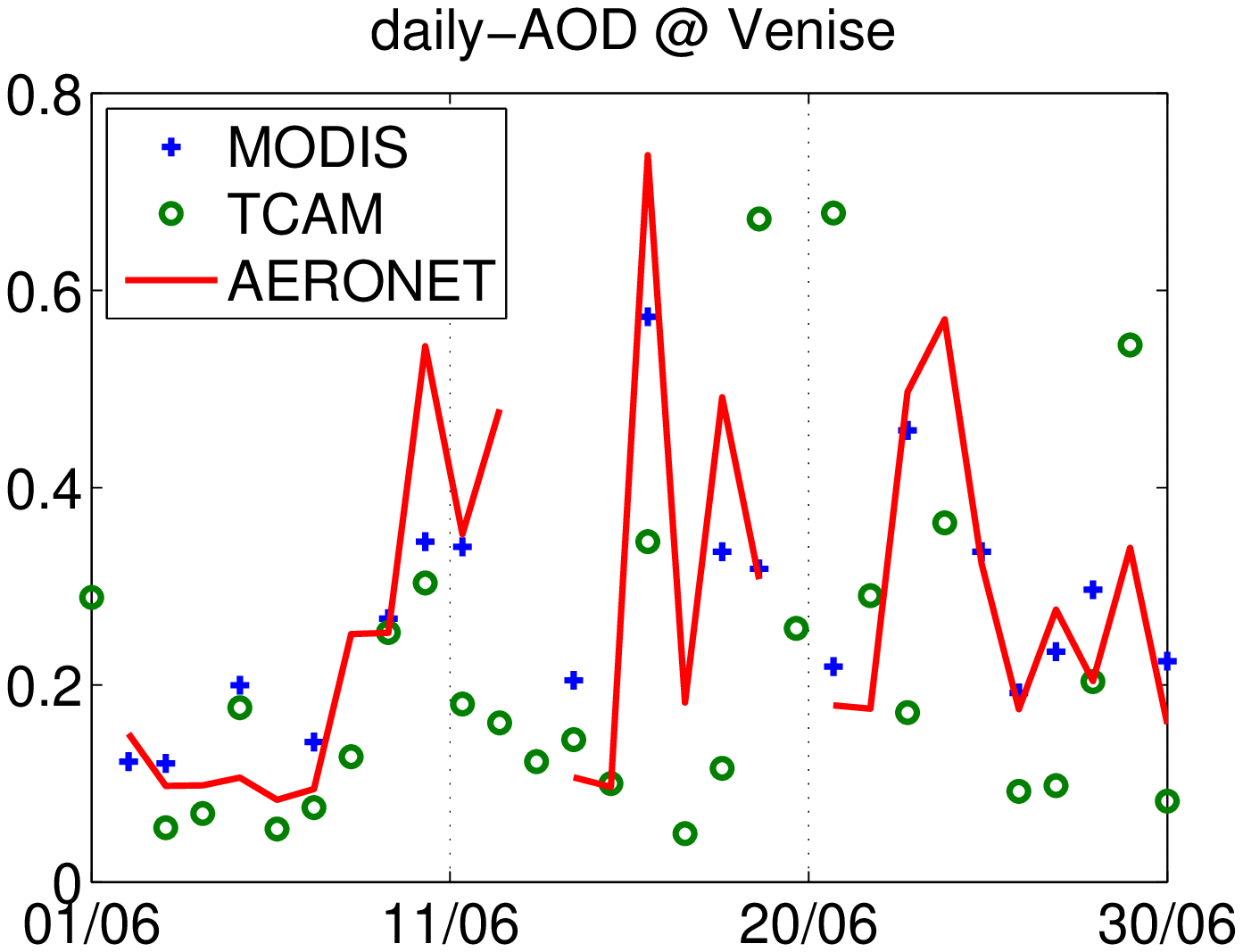}
\end{minipage}\hfill
\begin{minipage}{.5\linewidth}
\includegraphics*[width=7.cm] {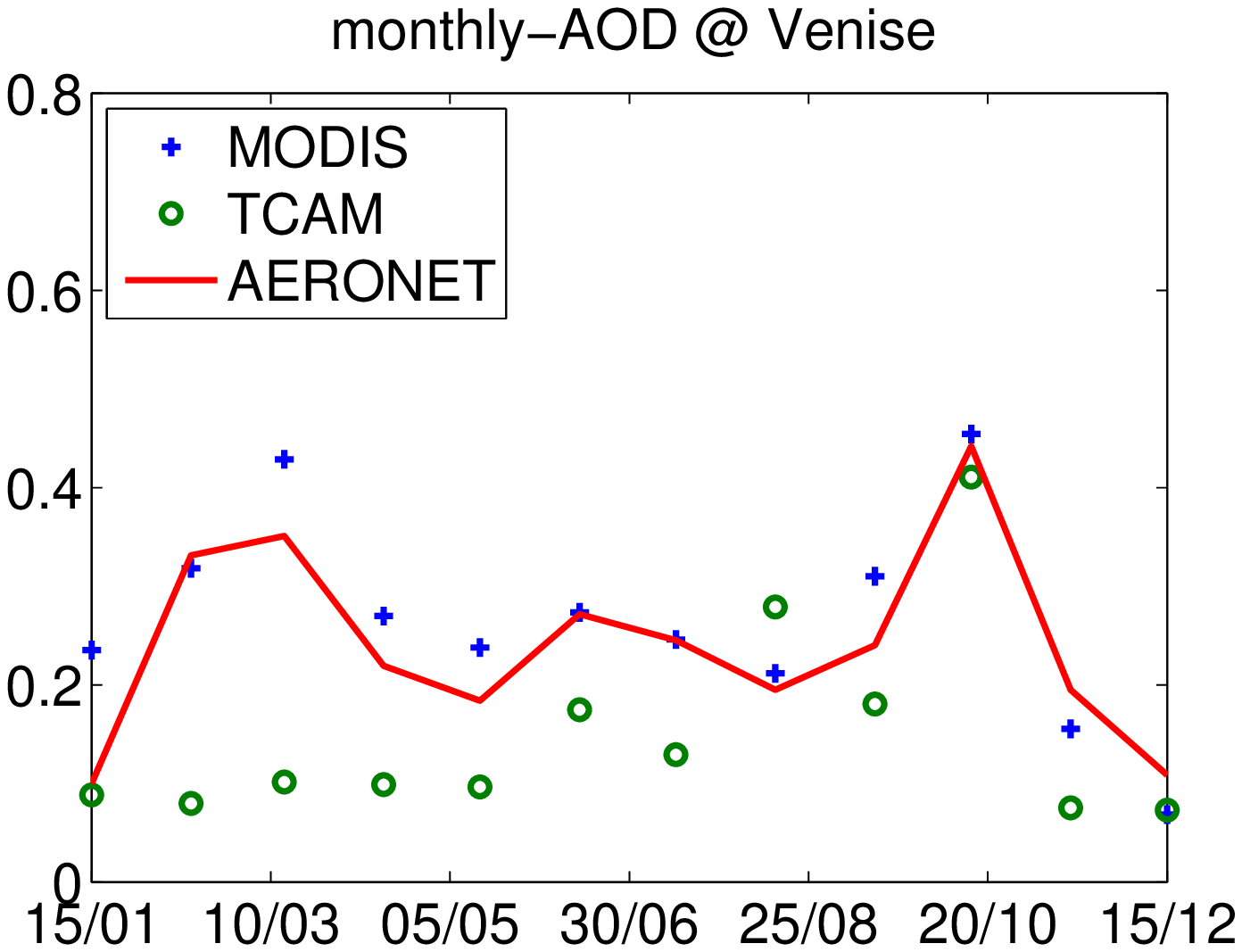}
\end{minipage}\hfill
\begin{minipage}{.5\linewidth}
\includegraphics*[width=7.cm] {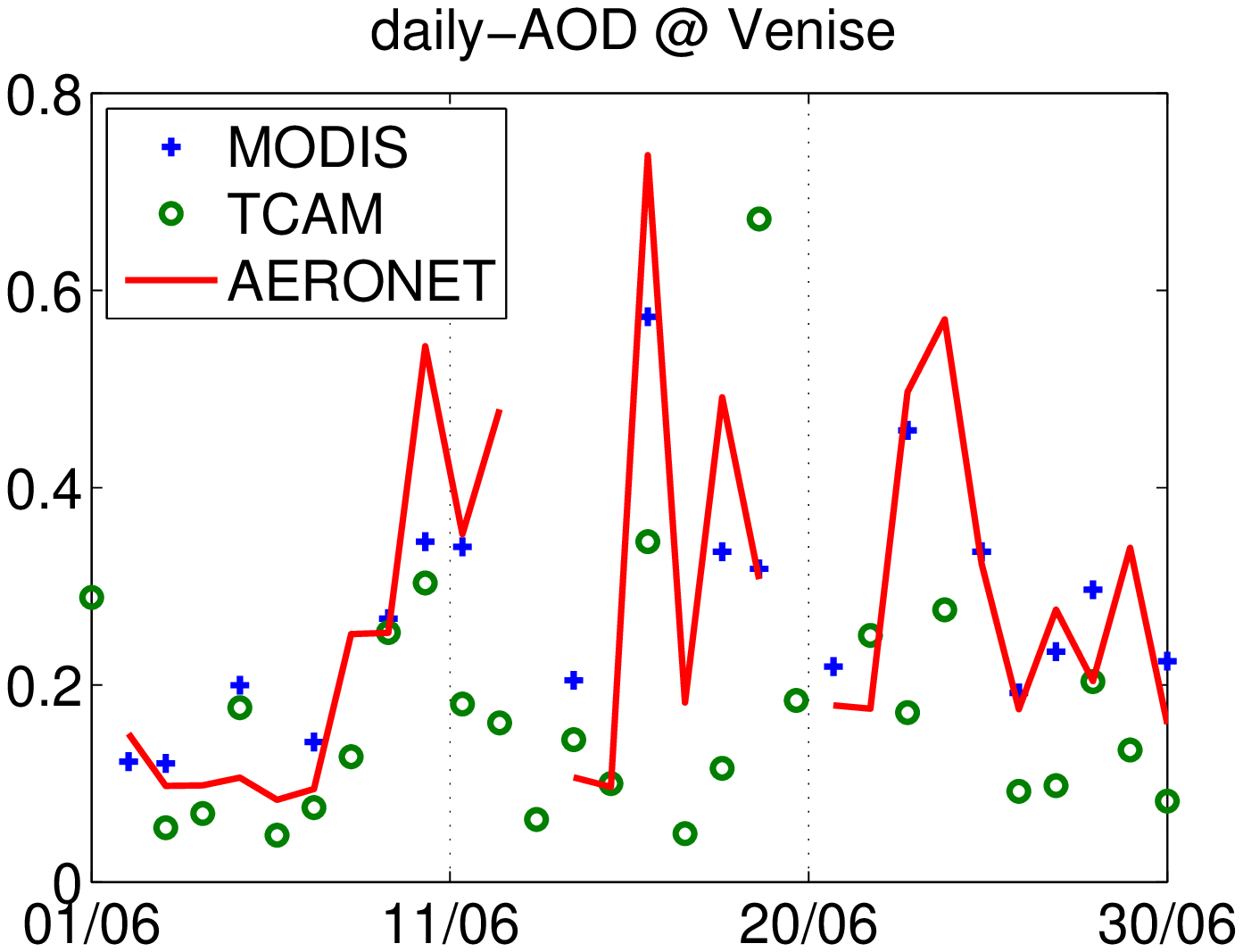}
\end{minipage}
\caption{ {\em AOD timeseries at  the QUITSAT cell next to AERONET site "Venise". Left panels refer to whole year 2004 while right panels are zooms on the month June 2004. Red lines for AERONET AOD, blue crosses for MODIS AOD, and green circles for TCAM AOD.  First row:   dry; Second row: wet; Third row: wet with cloud mask.}} \label {fig:AOD_timeseries}
\efi
Comparisons of spatial maps of AOD from satellite and AOD from the model can be useful for checking the ability of the model to reproduce spatial patterns. However, satellite AOD itself is affected by significant uncertainty: Minimum relative uncertainty on AOD is as large as 15\% \cite{levy07}. Thus, satellite information alone can not be used as a benchmark of model optical performance. 
The global network of sun-sky photometers AERONET is often used as a standard reference for AOD measurements \cite{dubo00}. There are 4 AERONET sites within Po-Valley, namely: Ispra, Modena, ISDGM-CNR, and Venise.  Version 2, level 2.0 AOD values at 870nm and 440nm from these sites have been used for estimating  AERONET AOD at 550 nm.    Angstr\"om  power law has been assumed to hold in this spectral range, as it is done in \cite{hodz04}

Comparisons of spatial maps of AOD from satellite and AOD from the model can be useful for checking the ability of the model to reproduce spatial patterns. However, satellite AOD itself is affected by significant uncertainty: Minimum relative uncertainty on AOD is as large as 15\% \cite{levy07}. Thus, satellite information alone can not be used as a benchmark of model optical performance. 
The global network of sun-sky photometers AERONET is often used as a standard reference for AOD measurements \cite{dubo00}. There are 4 AERONET sites within Po-Valley, namely: Ispra, Modena, ISDGM-CNR, and Venise.  Version 2, level 2.0 AOD values at 870nm and 440nm from these sites have been used for estimating  AERONET AOD at 550 nm.    Angstr\"om  power law has been assumed to hold in this spectral range, as it is done in \cite{hodz04}.

Comparisons of AOD daily mean values at one of the Po-Valley AERONET sites are presented  in \Fig {fig:AOD_timeseries}. Here, daily mean values correspond to an average over the times (within 60 min.) of both Terra and Aqua satellite overpasses. The Venise site has been selected for its representativeness: it is located at the east end of Po-Valley over an offshore oceanographic tower.  Instead, both Ispra and Modena sites are within a few tenth km from mountains. Finally, ISDGM-CNR photometer is located within the city of Venice and is possibly affected by local pollution sources.

In \Fig {fig:AOD_timeseries} both  whole-2004 monthly-averaged timeseries and zooms on the month of June 2004 are shown. Satellite and AERONET information is identical in the three rows of the Figure. First of all, one notices that there is a fair agreement between MODIS and AERONET AOD, with a maximum monthly value of about 0.4 in October. For such a high AOD, MODIS expected accuracy is better than 30\% \cite{levy07}.

Passing to model outputs, in the first row, timeseries of  TCAM dry AOD are considered. Their values are systematically lower than both satellite and AERONET AOD by nearly an order of magnitude. The June-zoom shows however that the model follows the same trend of AERONET in the first half of the month: in particular, a seemingly exponential build-up of AOD is observed from June 6 to June 10. This time  interval corresponds to stable weather conditions after rains which have interested Po-Valley on the first days of the month. 

In the second row of \Fig {fig:AOD_timeseries},  TCAM wet AOD is displayed. There is an impressive model AOD increase with respect to the dry case. In correspondence of winter months, the model delivers AOD values which are even higher than AERONET ones. Furthermore, minimum values of TCAM AOD are now much closer to observations.

In the third row, the cloud masking mechanism illustrated in \Sect{sec:opt_module_cloudmask} has been employed. As a consequence, model AOD is nearly halved for the first part of year 2004 and the number of June AOD overestimations is reduced. The October peak is very well reproduced by both AERONET, MODIS, and TCAM.

\section{Conclusions}\label{sec:conclusion}
An optical module for  chemistry transport model TCAM has been developed and resulting AOD has been compared to observations over the Po-Valley (Italy) during year 2004. 

The optical module takes aerosol concentrations as an input and delivers AOD fields as an output. Main inorganics (nitrates, ammonium, sulfates) and organic chemical compounds are considered. Within TCAM, a detailed description of the aerosol size distribution is achieved through 10 size bins, whose position and size can vary during simulations. For the computation of AOD, aerosol particles are modeled as homogeneous externally mixed spheres and the effect of moisture is accounted for through a parametrization.

 Thanks to the detailed representation of granulometric classes implemented into TCAM, it is found that the largest contribution to AOD arises from inorganic compounds in the sub-micrometric range. This is fully consistent with comparable literature: Hodzic et. al. too found that over 89\% of AOD arises from the accumulation mode \cite{hodz04}; Furthermore, in the WRF-chem modeling exercise, 6 out of 8 size bins where centered on sub-micron values of the aerosol radius  \cite{chap09}.

  TCAM AOD has been compared to both MODIS AOD operational product and to AERONET AOD. The model is able to capture main spatial and temporal features of the observations. However, dry model output is up to one order of magnitude smaller than observations. Quantitative agreement is significantly improved by the parametrization scheme used for dealing with  moisture. 
  
  With respect to the issue of quantitive agreement between model and observations,  one should keep in mind that the final purpose of AOD computation from TCAM aerosol concentration is to able to assimilate satellite AOD into TCAM. Departures between satellite and model AOD represent the actual information which can hopefully improve model performance in terms of aerosol concentration fields. However, it is important to stress at this place that even satellite AOD data are affected by significant uncertainties. MODIS retrieval algorithm over land implies a relative uncertainty on AOD between 15\% and 100\%, increasing for lower values of AOD \cite{levy07}. Thus,  an assimilation procedure to come has to account for errors of both model and satellite AOD in a proper way. To the extent of the computation of the optical properties, it is important that the physical features of the chemical and transport model are accounted for faithfully. These features include both the microphysical aerosol parameters (cp. \Tab{tab:aerclasses_LUT}), the modeling of the extinction coefficient (cp. \Eq{extcoeff_def}), and the parametrization of moisture (cp. \Eq{growth_factor}).
The best possible implementation of these issues leads to the "true" model AOD and corresponding "true" departures  between satellite and model AOD. In no case, except the case of accidental agreement, such true departures can be null, since the model is known to fail in satisfactory reproduction of observed PM10 data.  
Assimilation of such true  departures into the model will lead to consistent corrections to the model fields.

The optical module   can be used for data assimilation of any satellite AOD into TCAM. In particular, the present results suggest that this operation can be meaningful already at  mesoscale and not only at  global and regional scale as already tested in the literature.  This  is in the spirit of QUITSAT project, whose aim is to bring together model outputs and observations for a better evaluation of air-quality related variables.

\section{Acknowledgements}
This research has been developed in the framework of the Pilot Project QUITSAT, contract no. I/035/06/0, sponsored and funded by the Italian Space Agency (ASI). The CETEMPS group at Universit\agr dell'Aquila has provided boundary conditions and meteorological fields for feeding TCAM simulations. The principal investigator of the AERONET site Venise is kindly acknowledged.

\bibliographystyle{elsart-harv}
\bibliography{biblio}

\end{document}

%% file: command.tex
\newcommand{\be}{\begin{equation}}
\newcommand{\ee}{\end{equation}}
\newcommand{\bea}{\begin{eqnarray}}
\newcommand{\bean}{\begin{eqnarray}\nonumber}
\newcommand{\eea}{\end{eqnarray}}
\newcommand{\Eq}[1]{Eq.\,(\ref{#1})}

\newcommand{\Sect}[1]{Sect.\,(\ref{#1})}

\newcommand{\Fig}[1]{Fig.\,\ref{#1}}
\newcommand{\Tab}[1]{Tab.\,\ref{#1}}
\newcommand{\bfi}{\begin{figure}}
\newcommand{\efi}{\end{figure}}
\newcommand{\bmi}{\begin{minipage}}
\newcommand{\emi}{\end{minipage}}

\newcommand{\agr}{\`a }

\renewcommand{\Re}{\mbox{\sl Re\,}}